\documentclass[12pt]{article}
\usepackage{a4wide,amssymb,cite,latexsym}
\pdfoutput=1

\usepackage{a4wide,amssymb,graphicx}
\usepackage{epsfig}
\usepackage[usenames,dvipsnames]{color}
\usepackage{slashed}
\usepackage{tikz}
\usepackage[compat=1.1.0]{tikz-feynman}
\usepackage{xcolor}
\usepackage{color}
\usepackage{relsize}
\usepackage{graphics}
\usepackage{epstopdf}
\usepackage{hyperref}
\usepackage{mathrsfs}
\usepackage{gensymb}
\usepackage{amssymb}
\usepackage{amsthm}
\usepackage{amsmath}
\usepackage{hyperref}

\newcommand{\be}{\begin{equation}}
\newcommand{\ee}{\end{equation}}
\newcommand{\bea}{\begin{eqnarray}}
\newcommand{\eea}{\end{eqnarray}}

\def\circa#1{\,\raise.3ex\hbox{$#1$\kern-.75em\lower1ex\hbox{$\sim$}}\,}

\newcommand{\ba}{\begin{array}{c}}
\newcommand{\bat}{\begin{array}{cc}}
\newcommand{\ea}{\end{array}}
\newcommand{\beqn}{\begin{eqnarray}}
\newcommand{\eeqn}{\end{eqnarray}}

\newcommand{\bi}{\begin{itemize}}
\newcommand{\ei}{\end{itemize}}

\newcommand{\lab}{\langle}
\newcommand{\rab}{\rangle}

\begin{document}

\begin{titlepage}

\rightline{CAU-THEP-19-02, LPT-Orsay-19-12, IFT-UAM/CSIC-19-35}

\begin{centering}
\vspace{1cm}
{\Large {\bf Vector SIMP dark matter with approximate \vspace{0.2cm}\\ custodial symmetry}} \\

\vspace{1cm}

{\bf  Soo-Min Choi$^{1,\dagger}$, 
 Hyun Min Lee$^{1,2,*}$, Yann Mambrini$^{3,\sharp}$  and Mathias Pierre$^{4,5,\natural}$ }
\\
\vspace{.5cm}

{\it $^1$Department of Physics, Chung-Ang University, Seoul 06974, Korea.} 
\vspace{0.2cm} \\
{\it $^2$School of Physics, Korea Institute for Advanced Study, Seoul 02455, Korea.}
\vspace{0.2cm}\\
{\it $^3$Laboratoire de Physique Th\'eorique (UMR8627), CNRS, Univ. Paris-Sud,   Universit\'e Paris-Saclay, 91405 Orsay, France. } 
\vspace{0.2cm} \\
{\it $^4$Instituto de F\'isica Te\'orica (IFT) UAM-CSIC,
Campus de Cantoblanco, 28049 Madrid, Spain.
}
\vspace{0.2cm} \\
{\it $^5$Departamento de F\'isica Te\'orica, Universidad Aut\'onoma de Madrid (UAM), Campus de Cantoblanco, 28049 Madrid, Spain.
}

\end{centering}
\vspace{0.5cm}

\begin{abstract}
We consider a novel scenario for Vector Strongly Interacting Massive Particle (VSIMP) dark matter with local $SU(2)_X\times U(1)_{Z'}$ symmetry in the dark sector. Similarly to the Standard Model (SM), after the dark symmetry is broken spontaneously by the VEVs of dark Higgs fields, the approximate custodial symmetry determines comparable but split masses for $SU(2)_X$ gauge bosons. In this model, we show that the $U(1)_{Z'}$-charged gauge boson of $SU(2)_X$ ($X_\pm$) becomes a natural candidate for SIMP dark matter, annihilating through $3\rightarrow 2$ or forbidden $2\rightarrow 2$ annihilations due to gauge self-interactions. On the other hand, the $U(1)_{Z'}$-neutral  gauge boson of $SU(2)_X$ achieves the kinetic equilibrium of dark matter through a gauge kinetic mixing between $U(1)_{Z'}$ and SM hypercharge. We present the parameter space for the correct relic density in our model and discuss in detail the current constraints and projections from colliders and direct detection experiments.  

\end{abstract}

\vspace{0.5cm}

\begin{flushleft}
$^\dagger$Email: soominchoi90@gmail.com\\
$^*$Email: hminlee@cau.ac.kr  \\
$^\sharp$Email: yann.mambrini@th.u-psud.fr \\
$^\natural$Email: mathias.pierre@uam.es
\end{flushleft}

\end{titlepage}

\section{Introduction}

Despite indirect evidences for the presence of dark matter in our Universe \cite{planck}, the nature of dark matter is still elusive. The absence of 
 signals in direct detection experiments like LUX \cite{LUX}, PANDAX \cite{PANDAX} or more recently XENON1T \cite{XENON} questions the WIMP (Weakly Interacting Massive Particle) paradigm.  The simplest extensions, involving minimal ingredients as Higgs-portal \cite{hp,Higgsportal,hp2a,hp2b,hp2}, $Z$-portal \cite{Zportal} or even 
 $Z'$-portal \cite{Zpportal}, etc, are already excluded for Beyond the Standard Model (BSM) scale  below $\sim 3$ TeV, once
 we combine cosmological, accelerators and direct detection constraints (see \cite{Arcadi:2017kky}
 for a recent review on the subject and \cite{gambit} for the recent global fits for Higgs-portal dark matter). 
 In this context, one needs to develop extensions of the Standard Model beyond the WIMP paradigm. 

One possibility for going beyond the WIMP paradigm is to 
 modify drastically the thermal history of dark matter, allowing for extremely feeble couplings between the visible and dark sectors.
 The correct relic abundance is then ensured through the freeze-in mechanism \cite{fimp, fimpmodels,fimpgravitino}.   
 Another possibility proposed recently is to allow for large self-interacting couplings generating $3 \rightarrow 2$ or $2 \rightarrow 2$ forbidden channels
 \cite{simp,z3model,simpG,fdm,simpmodels,vsimp}, which can then provide solutions to some of small-scale problems at galaxy scales \cite{small-scale,smallscale2,ss22,ss3}. These Strongly Interacting Massive Particle (SIMP) scenarios, naturally present in the case of non-abelian dark matter, open up
 a complete new range of parameter space that still needs to be explored. 
 
 We propose in this work to analyse in detail a simple $SU(2)_X \times U(1)_{Z'}$ extension of the Standard Model (SM), where the charged non-abelian vector boson $X_{\pm}$ is the dark matter candidate, while the $Z'_\mu$ boson plays a role of the portal between  the visible and 
 hidden sectors through its kinetic mixing with the SM hypercharge gauge boson $B_\mu$.
Our construction provides a novel and efficient mechanism for maintaining VSIMP dark matter in kinetic equilibrium during freeze-out, being consistent with the observed relic density.  
We will discuss the interplay between correct relic density  and kinetic equilibrium in constraining the parameter space, which can be tested at current and future experiments. 
 
Our paper is organized as follows. We describe the model in the first section and compute its spectrum in Section 3. Section 4 is devoted to the computation of dark matter
abundance through the $3 \rightarrow 2$ and $2 \rightarrow 2$ forbidden channels while a combined analysis including 
constraints from direct detection and accelerators searches is presented in Section 5. We then conclude.  There are two appendices dealing with general masses for gauge and Higgs bosons in the dark sector.

%intro

%SIMP and small-scale problems

%VSIMP and kinetic equilibrium

%plan

\section{Model}

We consider models with a local $SU(2)_X\times U(1)_{Z'}$ symmetry in the dark sector, ``dark" in the sense that the Standard Model particles are not charged under these transformations.
The dark Higgs sector is composed of a singlet scalar $S$ and a nontrivial representation scalar $H_X$ under $SU(2)_X$, with $U(1)_{Z'}$ charges given by $q_S$ and $q_{H_X}$, respectively. 
Without loss of generality, we choose $q_{H_X}=I_3$ where $I_3$ is the dark isospin of $H_X$.

The Lagrangian for the dark sector in our model is then given by
\be
{\cal L}= -\frac{1}{4} {\vec X}_{\mu\nu}\cdot {\vec X}^{\mu\nu}-\frac{1}{4} Z'_{\mu\nu} Z^{\prime \mu\nu}+ \mathcal{L}_{\rm scalar}+{\cal L}_{Z'-{\rm portal}}+ {\cal L}_{H-{\rm portal}}\, \label{dLag}
\ee
where the field strength tensors are ${\vec X}_{\mu\nu}=\partial_\mu {\vec X}_\nu-\partial_\nu {\vec X}_\mu+g_X ({\vec X}_\mu\times {\vec X}_\nu) $ and $Z'_{\mu\nu}=\partial_\mu Z'_\nu -\partial_\nu Z'_\mu$. The scalar part of the Lagrangian containing the SM Higgs doublet $H$ is given by
\bea\label{eq:Vhiggs}
\mathcal{L}_{\rm scalar}&=&  |D_\mu S|^2 +m^2_S |S|^2 -\lambda_S |S|^4 + |D_\mu H_X|^2  \nonumber \\
&&+m^2_{H_X} |H_X|^2 -\lambda_{H_X} |H_X|^4+{\tilde\lambda}_{H_X} (H_X^\dagger t_i H_X)(H_X^\dagger t_i H_X)+ V_{\rm mix}(S,H_X)  \nonumber \\
&&+|D_\mu H|^2+m^2_H |H|^2 -\lambda_H |H|^4
\eea
where the covariant derivatives for $S$ and $H_X$ are 
\be
D_\mu=\partial_\mu-i g_X {\vec t}\cdot {\vec X}_\mu-i g_{Z'} q_{Z'} Z'_\mu
\ee
with $t^i (i=1,2,3)$ being $SU(2)_X$ generators satisfying $[t^i,t^j]=i\epsilon^{ijk}t^k$, $ V_{\rm mix}$  the scalar potential including mixing quartic couplings between dark Higgs fields\footnote{ Notice that we can set ${\tilde \lambda}_{H_X}=0$ when $H_X$ is a doublet.}, and $q_{Z'}=q_S$ or $q_{H_X}$. 
We also introduce a gauge kinetic mixing between $Z'$ and hypercharge gauge bosons,
\bea
{\cal L}_{Z'{\rm- portal}} = -\frac{1}{2} \,\sin\xi\, Z'_{\mu\nu} B^{\mu\nu}.  \label{kinmix}
\eea
This mixing will play the role of the portal between the dark sector and the Standard Model for the thermalization process. 
For completeness, we allow for Higgs portal couplings between the dark scalars and the SM Higgs,
\bea
{\cal L}_{H{\rm- portal}} =  -\lambda_{S H} |S|^2 |H|^2-\lambda_{H_X H}|H_X|^2|H|^2.  \label{Higgsp}
\eea

Moreover, developing eq.~(\ref{dLag}), we extract  self-interacting interactions of dark gauge bosons \cite{vsimp},
\bea
{\cal L}_{\rm self} &=&-\frac{1}{2}g_X (\partial_\mu {\vec X}_\nu -\partial_\nu {\vec X}_\mu)\cdot ({\vec X}^\mu\times {\vec X}^\nu) -\frac{1}{4} g^2_X ({\vec X}_\mu\cdot {\vec X}^\mu)^2 \nonumber \\
&&+\frac{1}{4}g^2_X({\vec X}_\nu\cdot {\vec X}^\mu)({\vec X}_\mu\cdot {\vec X}^\nu)\, \nonumber \\
&\equiv &{\cal L}_3 +{\cal L}_4 \label{self}
\eea
with 
\bea
{\cal L}_3 &=&-i g_X\bigg[
\left(\partial^\mu X^\nu -\partial^\nu X^\mu\right)
 X^\dagger_\mu X_{3,\nu} -
\left(\partial^\mu X^{\nu\dagger} -\partial^\nu X^{\mu\dagger}\right)
 X_\mu X_{3,\nu} \nonumber \\
&& + X_\mu X^\dagger_\nu\left(\partial^\mu X^\nu_3 -\partial^\nu X^\mu_3\right)
\bigg],
 \nonumber \\
{\cal L}_4&=& -{g^2_X\over 2}\left[
\left(X^\dagger_\mu X^\mu\right)^2 - X^\dagger_\mu X^{\mu\dagger}
X_\nu X^\nu \right]
- g^2_X\,\left(
X_\mu^\dagger X^\mu X_{\nu,3} X^\nu_3 - X^\dagger_\mu X^\mu_3 X_\nu X^\nu_3
\right)
\eea
where $X_\mu\equiv (X_{1,\mu}+i X_{2,\mu})/\sqrt{2}$ and its complex conjugate, $X^\dagger_\mu\equiv (X_{1,\mu}-i X_{2,\mu})/\sqrt{2}$. 
See also Ref.~\cite{NA-dm} for some of non-abelian dark matter candidates.

After $SU(2)_X\times U(1)_{Z'}$ is broken by the VEVs of dark scalars, the dark gauge bosons have nonzero masses, and $X_{1,2,\mu}$ gauge bosons are combined to be a complex gauge boson $X_\mu$ with nonzero charge under $U(1)_{Z'}$. Thus, $X_{1,2,\mu}$ gauge bosons, if the lightest particle with $U(1)_{Z'}$ charge in the dark sector,  can be a dark matter candidate.
Indeed, as the Standard Model is neutral under $U(1)_{Z'}$, no decay modes of $X_\mu$ into the visible sector are allowed due to the DM $Z_2$ symmetry after spontaneous breaking of $U(1)_{Z'}$. On the other hand, as it happens in the Standard Model with $W^{\pm}$ and $Z^0$, $X_{3\mu}$ has a different mass from  $X_{1,2\mu}$ due to the VEV of  $H_X$ charged under the $U(1)_{Z'}$, and it can couple to the SM particles through the dark Weinberg angle in combination with the gauge kinetic mixing, as it will be discussed later. Our situation is a mirror case of the Standard Model, except for the presence of an extra-singlet $S$ avoiding the presence of a massless gauge boson.

\section{Dark gauge boson masses}

In this section, we discuss the approximate custodial symmetry in the $SU(2)_X$ gauge sector with general dark Higgs representations and show the effect of the $Z'$ gauge boson on the mass splitting between dark-neutral and charged gauge bosons.

\subsection{Dark custodial symmetry and its breaking}

We considered the expansions of dark Higgs fields about nonzero VEVs by a singlet $S=\frac{1}{\sqrt{2}}\,(v_S+s)$, and a Higgs field $H_X$ in several representation of $SU(2)_X$: a doublet $\Phi=(0,\frac{1}{\sqrt{2}}(v_\Phi+h_\Phi))^T$, a triplet $T=(h^{++},0,\frac{1}{\sqrt{2}}(v_T+h_T))^T$ or  a quadruplet $Q_4=(h^{(3)},h^{(2)},0,\frac{1}{\sqrt{2}}(v_{Q_4}+h_{Q_4}))^T$ or a quintuplet $Q_5=(h^{(4)},h^{(3)},h^{(2)},0,\frac{1}{\sqrt{2}}(v_{Q_5}+h_{Q_5}))^T$, in unitary gauge. 
We present general masses for dark gauge bosons in appendix A, 
while the expression for the dark Higgs masses and the mixing between the dark Higgs 
and the SM Higgs are given in appendix B.

First, due to the VEV of the dark Higgs $H_X$ in a nontrivial representation of $SU(2)_X$, such as $\Phi$, $T$, $Q_4$ or $Q_5$, masses of dark-charged gauge bosons, $X_\mu, X^\dagger_\mu$,  are given by
\be
m^2_X= \frac{1}{2} g^2_X I v^2_I\, \label{mass10}
\ee
with $I=\frac{1}{2}, 1, \frac{3}{2}, 2$ respectively. 
For a vanishing $Z'$ charge of the dark Higgs $H_X$ or in the limit of a vanishing $g_{Z'}$, the dark-neutral gauge boson has mass
\bea
m^2_{X_3,0}= g^2_X I^2 v^2_I, 
\eea
leading to the mass relations due to {\it dark custodial symmetry},
\bea
\frac{m^2_X}{m^2_{X_3,0}}= \frac{1}{2I}.
\eea

\noindent
Notice that we recover the $M_Z=M_W$ relation of the Standard Model with a Higgs doublet ($I=\frac{1}{2}$) in the case of a null Weinberg angle. 
As a result, if $I>\frac{1}{2}$, the dark-charged gauge boson $X$ can be the lightest gauge boson in the dark sector, becoming a candidate for non-abelian dark matter. On the other hand, once taking into consideration the VEV of $S$, $Z'$ gauge boson can decouple from the $X$ spectrum due to the contribution of $v_S$ to its mass.
To maintain the thermal equilibrium of $X_\mu$ with the Standard Model bath, one needs to consider first its coupling to $Z'$.

Indeed, in order to communicate between non-abelian dark matter and SM by renormalizable couplings, it is sufficient to consider the dark Higgs $H_X$ with nonzero $Z'$ charge and take a sizable $g_{Z'}$.
Then,  the mass matrix for neutral gauge bosons in the basis ($Z'_\mu$,  $X_{3\mu}$), receives a correction term violating the dark custodial symmetry, as follows,
\bea
M^2_{2\times 2}=m^2_{X_3}\left(\begin{array}{cc} \alpha s^2_X  & -s_X c_X \\   -s_X c_X  & c^2_X  \end{array} \right) \label{mass220}
\eea
where $m^2_{X_3}\equiv (g^2_X+g^2_{Z'}) I^2 v^2_I=m^2_{X_3,0}+g^2_{Z'} I^2 v^2_I$, 
$c_X\equiv \cos\theta_X$ and $s_X\equiv \sin\theta_X$, with $\sin\theta_X=g_{Z'}/\sqrt{g^2_X+g^2_{Z'}}$, and 
\bea
\alpha\equiv 1+\frac{q^2_S v^2_S}{I^2 v^2_I}. 
\eea
Unlike in the SM, there exists a singlet scalar $S$ contributing to the $Z'$ mass.
The most general dark gauge boson masses with VEVs of Higgs fields in arbitrary representations are given in the appendix A.

In the absence of the gauge kinetic mixing, the mass matrix (\ref{mass220}) can be diagonalized explicitly by introducing a dark Weinberg angle as in the SM. 
Performing a rotation of dark gauge fields to mass eigenstates\footnote{In the presence of a gauge kinetic mixing between $Z'$ and hypercharge gauge boson,  mass eigenstates as well as mass eigenvalues of neutral gauge bosons including the $Z$-boson are modified, but we consider the case where mass corrections are negligible but new interactions of extra gauge bosons to the SM are kept in the leading order in the gauge kinetic mixing parameter, as will be discussed in the next section. }, ${\tilde Z}'_\mu, {\tilde X}_{3\mu}$,  as
\bea
\left(\begin{array}{c}  Z'_\mu \\ X_{3\mu} \end{array} \right)= \left(\begin{array}{cc} \cos\theta'_X & -\sin\theta'_X \\   \sin\theta'_X & \cos\theta'_X \end{array} \right) \left(\begin{array}{c}  {\tilde Z}'_\mu \\ {\tilde X}_{3\mu} \end{array} \right)
\eea
with
\bea
\tan(2\, \theta'_X)= \frac{2c_X s_X}{c^2_X-\alpha\, s^2_X}, \label{gaugemix0}
\eea
we obtain the mass eigenvalues for dark gauge bosons,
\bea
m^2_{ {\tilde Z}'} &=&m^2_{X_3}c^2_X  (1-\cot\theta'_X\, \tan\theta_X),  
\label{mass20} \\
m^2_{{\tilde X}_{3}} &=& m^2_{X_3}c^2_X(1+\tan\theta'_X\, \tan\theta_X). \label{mass30}
\eea
Therefore, from the results in eqs.~(\ref{mass10})-(\ref{mass30}),  we can keep the hierarchy of masses,
\bea
m^2_X< m^2_{{\tilde X}_3} < m^2_{{\tilde Z}'}\, ,
\eea
as far as
\bea
\tan\theta'_X<0, \quad \quad |\tan\theta'_X|< \frac{1}{2\tan\theta_X}. 
\eea
In this case, the dark charged gauge boson is still the lightest gauge boson in the dark sector, so that the $2\rightarrow 2$ annihilation of $X_\mu, X^\dagger_\mu$ is forbidden while  $3\rightarrow 2$ processes with gauge self-interactions become dominant for determining the relic density of $X_\mu, X^\dagger_\mu$.

We note that if $m_X>m_{{\tilde X}_3}$, the $2\rightarrow 2$ annihilation of $X_\mu, X^\dagger_\mu$ into a ${\tilde X}_3$ pair is open, dominating the relic abundance calculation and leading to an interesting possibility for WIMP dark matter. However, in this work, we are interested in the production of light dark matter below sub-GeV scale, so henceforth we focus on the case with $m_X<m_{{\tilde X}_3}$.

Due to the mixing between $Z'_\mu$ and $X_{3\mu}$, self-interactions  for $SU(2)_X$ gauge bosons in  eq.~(\ref{self}) become, in the basis of mass eigenstates,
\bea
{\cal L}_{\rm \rm self}={\cal L}_3 +{\cal L}_4
\eea
with
\bea
{\cal L}_3 &=& -i g_X\cos\theta'_X \bigg[
\left(\partial^\mu X^\nu -\partial^\nu X^\mu\right)
 X^\dagger_\mu {\tilde X}_{3,\nu} -
\left(\partial^\mu X^{\nu\dagger} -\partial^\nu X^{\mu\dagger}\right)
 X_\mu {\tilde X}_{3,\nu} \nonumber \\
&& + X_\mu X^\dagger_\nu\left(\partial^\mu {\tilde X}^\nu_3 -\partial^\nu {\tilde X}^\mu_3\right)
\bigg] \nonumber \\
&&-i g_X\sin\theta'_X \bigg[
\left(\partial^\mu X^\nu -\partial^\nu X^\mu\right)
 X^\dagger_\mu {\tilde Z}'_{\nu} -
\left(\partial^\mu X^{\nu\dagger} -\partial^\nu X^{\mu\dagger}\right)
 X_\mu {\tilde Z}'_{\nu} \nonumber \\
&& + X_\mu X^\dagger_\nu\left(\partial^\mu {\tilde Z}^{\prime\nu} -\partial^\nu {\tilde Z}^{\prime\mu}\right)\bigg], \\
{\cal L}_4 &=& -{g^2_X\over 2}\left[
\left(X^\dagger_\mu X^\mu\right)^2 - X^\dagger_\mu X^{\mu\dagger}
X_\nu X^\nu \right] \nonumber \\
&&-g^2_X \cos^2\theta'_X\,\left(
X_\mu^\dagger X^\mu {\tilde X}_{\nu,3} {\tilde X}^\nu_3 - X^\dagger_\mu {\tilde X}^\mu_3 X_\nu {\tilde X}^\nu_3 \right) \nonumber \\
&&-g^2_X \sin^2\theta'_X\,\left(
X_\mu^\dagger X^\mu {\tilde Z}'_{\nu} {\tilde Z}^{\prime\nu} - X^\dagger_\mu {\tilde Z}^{\prime\mu} X_\nu {\tilde Z}^{\prime\nu} \right) \nonumber \\
&& -g^2_X\sin\theta'_X \cos\theta'_X \,\left(
2X_\mu^\dagger X^\mu {\tilde X}_{3,\nu} {\tilde Z}^{\prime\nu} - X^\dagger_\mu {\tilde X}^{\mu}_3 X_\nu {\tilde Z}^{\prime\nu}  - X^\dagger_\mu {\tilde Z}^{\prime\mu} X_\nu {\tilde X}^{\nu}_3 \right). 
\eea
Moreover, in the basis of mass eigenstates for dark gauge bosons, for instance, from eq.~(\ref{higgs-gauge}) in the case of the triplet dark Higgs, we can also obtain the interactions between dark Higgs and gauge bosons. 
Concerning the self-interacting processes for dark-charged gauge bosons, we can use the above interactions by ignoring the mixing between dark Higgs and SM Higgs bosons and the mixings between dark-neutral gauge bosons and SM neutral gauge bosons.

\subsection{Split dark gauge bosons}

In the limit of $\alpha\gg 1$, the $Z'$ boson decouples from $X$ and $\tilde X_3$. From eq.~(\ref{gaugemix0}), we get $\tan\theta'_X\approx -\frac{1}{\alpha\tan\theta_X}$, leading to the approximate gauge boson masses,
\bea
m^2_{ {\tilde Z}'} &\approx& g^2_X I^2 v^2_I \Big( 1+\alpha \tan^2\theta_X\Big), \\
m^2_{{\tilde X}_{3}} &\approx& g^2_X  I^2 v^2_I \Big( 1-\frac{1}{\alpha}\Big),
\eea
leading to
\bea
m^2_{{\tilde X}_{3}}&\approx & 2I m^2_X \Big(1-\frac{1}{\alpha} \Big). \label{x3xmass}
\eea
Then, in order for the dark-charged gauge boson $X$ to be a viable dark matter candidate, we require $m_{{\tilde X}_{3}} >m_X$, which means $I>\frac{1}{2}$, that is, at least a triplet Higgs field with nonzero VEV must be introduced.
Moreover, in order for SIMP processes $XXX\rightarrow X{\tilde X}_3$ 
(which should be the dominant one in the relic abundance calculation) to be kinematically allowed, we also require $m_{{\tilde X}_{3}}<2m_X$, which means $I\leq 2$.
Thus, we need $m_X<m_{{\tilde X}_{3}}<2m_X $ in order to realize a viable SIMP dark matter in our model.

From eq.~(\ref{x3xmass}), we get the general expression for the mass splitting parameter in the limit of large $\alpha$:
\bea
\Delta\equiv \frac{m_{{\tilde X}_{3}}}{m_X}-1&\approx & \sqrt{2I}\sqrt{1-\frac{1}{\alpha}}-1 \nonumber \\
&\approx& \sqrt{2I}-1 -\frac{\sqrt{2I}}{2\alpha}, \label{split}
\eea
which will be relevant for forbidden channels in the later section.
As a result, the mass relations between dark gauge bosons are dictated by dark custodial symmetry at the leading order, but up to small corrections due to the mixing with $Z'$. 
For instance, we get $\Delta\approx \sqrt{2}-1, \sqrt{3}-1, 1$ for $I=1, \frac{3}{2}, 2$, respectively. 
Then, for the triplet Higgs case with $\Delta\approx \sqrt{2}-1$, we need to consider the $2\rightarrow 2$ forbidden channels for dark matter annihilations. On the other hand, for quadruplet with $I=\frac{3}{2}$ or quintuplet with $I=2$, the mass splitting being too large, it is
the $XXX \rightarrow X  {\tilde X}_3$ process which is dominant. 
For general dark Higgs VEVs, we can make $\Delta$ to vary continuously from $\Delta\approx 0$ to $\Delta\approx 2$. 
We keep this fact in mind for scanning the parameter space for the correct relic density with general $X_3$ masses in the next section.

\subsection{Degenerate dark gauge bosons}

A quick look at eqs.~(\ref{mass10}) and (\ref{mass30}) shows that there is a possibility of having degenerate mass terms in the dark sector,   $m_{{\tilde X}_{3}} \approx m_X$ provided that $\tan\theta'_X\approx -\frac{1}{2I}(2I-1) \cot\theta_X$ which gives, once combined with eq.~(\ref{gaugemix0}) :  $\tan\theta_X\approx \frac{1}{\sqrt{2I}}\sqrt{\frac{1}{\alpha-(2I)/(2I-1)}}$. Then, we obtain the abelian dark gauge boson mass as
\bea
m^2_{ {\tilde Z}'} \approx 2m^2_X (1+\tan^2\theta_X).
\eea
In this case,  in the limit of $\tan\theta_X\gg 1$ or  $g_{Z'}\gg g_X$ for $\alpha\approx \frac{2I}{2I-1}>0$, we can have ${\tilde Z}'$ decoupled. This case is possible for a nonzero VEV of the Higgs field with $I=1, \frac{3}{2}, 2$.
Then, all the non-abelian gauge bosons of $SU(2)$ can participate in the full $3\rightarrow 2$ processes, without Boltzmann suppression. 
But, in this case, the mixing angle between dark neutral gauge bosons become suppressed due to $|\tan\theta'_X|\ll 1$. Therefore, DM-SM elastic scattering through the kinetic mixing would be suppressed, so we need to rely on Higgs portal coupling for kinetic equilibrium. 

If one considers multiple Higgs fields with different isospins, in particular, doublet $\Phi$ and triplet $T$, from eq.~(\ref{genmassdiff}), we get the approximate mass difference in the limit of a large ${\tilde Z}'$ mass : 
\bea
m^2_{{\tilde X}_3}-m^2_X\approx \frac{1}{2} g^2_X v^2_T - \frac{1}{\beta}\, g^2_X \Big(\frac{1}{4}v^2_\Phi+v^2_T\Big)
\eea
with
\bea
\beta= 1+\frac{q^2_S v^2_S}{\frac{1}{2} v^2_\Phi+ v^2_T}\gg 1.
\eea
Then, we can tune dark Higgs VEVs, $v_\Phi$ and $v_T$ such that $m^2_{{\tilde X}_3}\approx m^2_X$.  In this case, as discussed in the Appendix, the mixing angle between dark neutral gauge bosons becomes 
\bea
\tan\theta'_X\approx -\frac{1}{\beta\tan\theta_X}. 
\eea
Therefore, even in this case, the mixing angle in the dark gauge sector is small so Higgs portal coupling would be more relevant for kinetic equilibrium as in the previous case.

\section{Dark matter annihilations with self-interactions}

We consider the Boltzmann equations for determining the dark matter relic density mainly by the DM self-interactions, in particular, by taking $3\rightarrow 2$ channels and forbidden $2\rightarrow 2$ channels  to be dominant.   We also present the results on DM self-scattering cross sections and impose current astrophysical limits on them together with the condition for a correct relic density.

\subsection{Boltzmann equations}

For $CP$ conserving interactions of dark matter, we get $n_{X^+} = n_{X^-}$.
Then, the total DM number density, $n_X=n_{X^+}+n_{X^-}$, is governed by the following Boltzmann equation,
\bea
\dot{n}_X + 3H n_X & =& -\lab \sigma v_{\rm rel} \rab_{2\rightarrow 2,v}\, (n_X^2 - n_{X,{\rm eq}}^2)  - \lab \sigma v_{\rm rel} \rab_{3\rightarrow 2}\, (n_X^3 - n_Xn_{X,{\rm eq}}^2) \nonumber \\
&&+ \lab \sigma v_{\rm rel} \rab_{2\rightarrow 2,d} \, n_{\tilde{X}_3,{\rm eq}}^2\left(
1-\frac{n^2_X}{n^2_{X,{\rm eq}}} \right)
\label{Boltz}
\eea
where
\bea
\lab \sigma v_{\rm rel} \rab_{2\rightarrow 2,v}&\equiv&\frac{1}{2}\lab \sigma v_{\rm rel} \rab_{X_+X_-\rightarrow f{\bar f}}, \\
\lab \sigma v_{\rm rel} \rab_{3\rightarrow 2}&\equiv&\frac{1}{2}\lab \sigma v^2_{\rm rel} \rab_{X_+X_+X_-\rightarrow X_+\tilde{X}_3}, \\
\lab \sigma v_{\rm rel} \rab_{2\rightarrow 2,d}&\equiv & 2\lab \sigma v_{\rm rel} \rab_{\tilde{X}_3\tilde{X}_3\rightarrow X_+X_-}.
\eea
Here, in the last line on right in eq.~(\ref{Boltz}), we have already used the detailed balance condition for forbidden channels to obtain the cross section for forbidden $2\rightarrow 2$ annihilation, $ X_+ X_-\rightarrow \tilde{X}_3 \tilde{X}_3$, as follows, 
\bea
\lab \sigma v_{\rm rel} \rab_{X_+X_-\rightarrow \tilde{X}_3\tilde{X}_3}=  \frac{4n_{\tilde{X}_3,{\rm eq}}^2}{n_{X,{\rm eq}}^2}\, \lab \sigma v_{\rm rel} \rab_{\tilde{X}_3\tilde{X}_3\rightarrow X_+X_-}.
\eea
Here, we assumed that ${\tilde X}_3$ keeps in thermal equilibrium with the SM during freeze-out, and the equilibrium abundances for $X$ and ${\tilde X}_3$ are given by
\be
Y_{X,\rm eq} =\frac{n_{X,{\rm eq}}}{s}= \frac{45x^2}{2g_{*s}\pi^4}\, K_2(x),\qquad Y_{\tilde{X}_3,\rm eq}=\frac{n_{{\tilde X}_3,{\rm eq}}}{s}=  \frac{45 m_{\tilde{X}_3}^2x^2}{4g_{*s}\pi^4 m_X^2}\, K_2\Big(\frac{m_{\tilde{X}_3}x}{m_X}\Big)
\ee
with $x\equiv m_X/T$. 
We have also included the SM $2\rightarrow 2$ annihilations, $X_+X_-\rightarrow f{\bar f}$, in the Boltzmann equation (\ref{Boltz}), as will be discussed in the next section.

\subsection{$3\rightarrow 2$ annihilations}

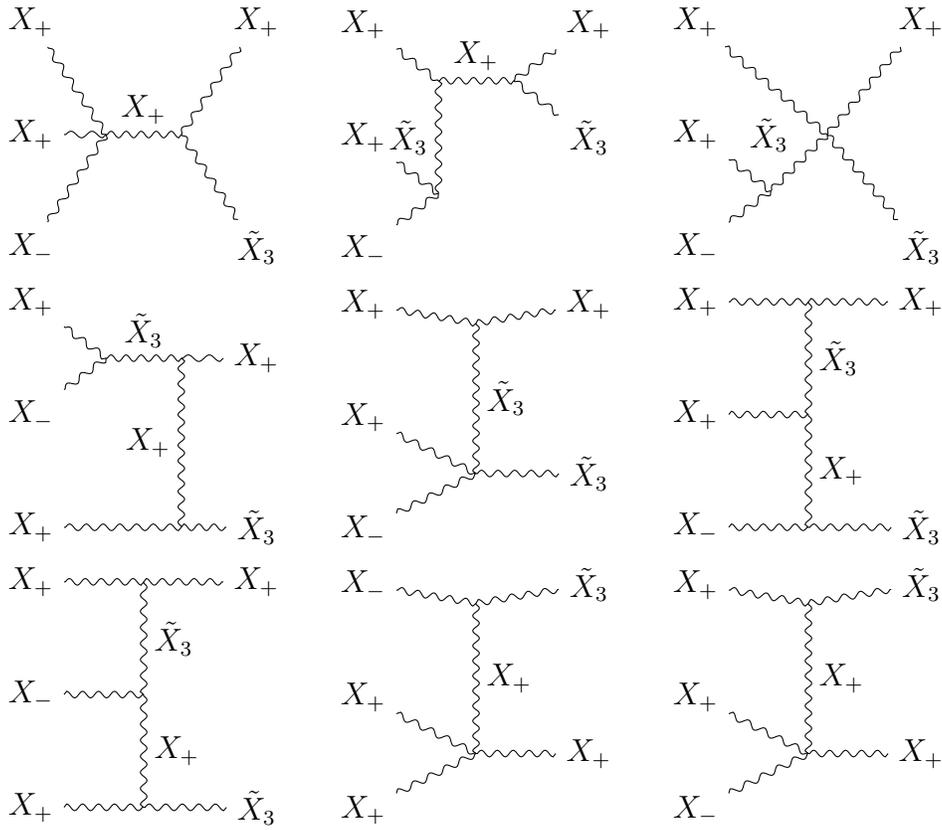
\begin{figure}
  \begin{center}
    \begin{tikzpicture}
\begin{feynman}
    \vertex (i1) {$X_+$};
    \vertex at ($(i1) + (0.0cm, -1.5cm)$) (i2) {$X_+$};
    \vertex at ($(i1) + (0.0cm, -3cm)$) (i3) {$X_-$};
    \vertex at ($(i1) + (3cm, +0.0cm)$) (f1) {$X_+$};
    \vertex at ($(i1) + (3cm, -3cm)$) (f2) {$\tilde{X}_3$};
    \vertex at ($(i1) + (1cm, -1.5cm)$) (m1);
    \vertex at ($(i1) + (2cm, -1.5cm)$) (m2);
    \diagram {
      {(i1),(i2),(i3)} -- [photon] (m1) -- [photon, edge label = $X_+$] (m2) -- [photon] {(f1),(f2)},      
    };
\end{feynman}
\end{tikzpicture}
\quad
%2
\begin{tikzpicture}
\begin{feynman}
    \vertex (i1) {$X_+$};
    \vertex at ($(i1) + (0.0cm, -1.5cm)$) (i2) {$X_+$};
    \vertex at ($(i1) + (0.0cm, -3cm)$) (i3) {$X_-$};
    \vertex at ($(i1) + (3cm, -0.0cm)$) (f1) {$X_+$};
    \vertex at ($(i1) + (3cm, -1.5cm)$) (f2) {$\tilde{X}_3$};
    \vertex at ($(i1) + (1cm, -2.25cm)$) (m1);
    \vertex at ($(i1) + (1cm, -0.75cm)$) (m2);
    \vertex at ($(i1) + (2cm, -0.75cm)$) (m3);
    \diagram {
      {(i2),(i3)} -- [photon] (m1) -- [photon, edge label = $\tilde{X}_3$] (m2) -- [photon] (i1),
      (m2) -- [photon, edge label = $X_+$] (m3) -- [photon] {(f1),(f2)}     
    };
\end{feynman}
\end{tikzpicture}
\quad
%3
\begin{tikzpicture}
\begin{feynman}
    \vertex (i1) {$X_+$};
    \vertex at ($(i1) + (0.0cm, -1.5cm)$) (i2) {$X_+$};
    \vertex at ($(i1) + (0.0cm, -3cm)$) (i3) {$X_-$};
    \vertex at ($(i1) + (3cm, 0cm)$) (f1) {$X_+$};
    \vertex at ($(i1) + (3cm, -3cm)$) (f2) {$\tilde{X}_3$};
    \vertex at ($(i1) + (1cm, -2.25cm)$) (m1);
    \vertex at ($(i1) + (1.75cm, -1.5cm)$) (m2);
    \diagram {
      {(i2),(i3)} -- [photon] (m1) -- [photon, edge label = $\tilde{X}_3$] (m2) -- [photon] {(i1),(f1),(f2)},
    };
\end{feynman}
\end{tikzpicture}
\\
%4
\begin{tikzpicture}
\begin{feynman}
    \vertex (i1) {$X_+$};
    \vertex at ($(i1) + (0.0cm, -1.5cm)$) (i2) {$X_-$};
    \vertex at ($(i1) + (0.0cm, -3cm)$) (i3) {$X_+$};
    \vertex at ($(i1) + (3cm, -0.75cm)$) (f1) {$X_+$};
    \vertex at ($(i1) + (3cm, -3cm)$) (f2) {$\tilde{X}_3$};
    \vertex at ($(i1) + (1cm, -0.75cm)$) (m1);
    \vertex at ($(i1) + (2cm, -0.75cm)$) (m2);
    \vertex at ($(i1) + (2cm, -3cm)$) (m3);
    \diagram {
      {(i1),(i2)} -- [photon] (m1) -- [photon, edge label = $\tilde{X}_3$] (m2) -- [photon] {(f1)},
      {(i3),(f2)} -- [photon] (m3) -- [photon, edge label = $X_+$] (m2),    
    };
\end{feynman}
\end{tikzpicture}
\quad
%5
\begin{tikzpicture}
\begin{feynman}
    \vertex (i1) {$X_+$};
    \vertex at ($(i1) + (0.0cm, -1.5cm)$) (i2) {$X_+$};
    \vertex at ($(i1) + (0.0cm, -3cm)$) (i3) {$X_-$};
    \vertex at ($(i1) + (3cm, -0cm)$) (f1) {$X_+$};
    \vertex at ($(i1) + (3cm, -2.25cm)$) (f2) {$\tilde{X}_3$};
    \vertex at ($(i1) + (1.5cm, -0.25cm)$) (m1);
    \vertex at ($(i1) + (1.5cm, -2.25cm)$) (m2);
    \diagram {
      {(i1),(f1)} -- [photon] (m1) -- [photon, edge label = $\tilde{X}_3$] (m2) -- [photon] {(i2),(i3),(f2)},      
    };
\end{feynman}
\end{tikzpicture}
\quad
%6
\begin{tikzpicture}
\begin{feynman}
    \vertex (i1) {$X_+$};
    \vertex at ($(i1) + (0.0cm, -1.5cm)$) (i2) {$X_+$};
    \vertex at ($(i1) + (0.0cm, -3cm)$) (i3) {$X_-$};
    \vertex at ($(i1) + (3cm, -0cm)$) (f1) {$X_+$};
    \vertex at ($(i1) + (3cm, -3cm)$) (f2) {$\tilde{X}_3$};
    \vertex at ($(i1) + (1.5cm, -0cm)$) (m1);
    \vertex at ($(i1) + (1.5cm, -1.5cm)$) (m2);
    \vertex at ($(i1) + (1.5cm, -3cm)$) (m3);
    \diagram {
      {(i1),(f1)} -- [photon] (m1) -- [photon, edge label = $\tilde{X}_3$] (m2) -- [photon, edge label = $X_+$] (m3) -- [photon] {(i3),(f2)},
      (m2) -- [photon] (i2)      
    };
\end{feynman}
\end{tikzpicture}
\\
%7
\begin{tikzpicture}
\begin{feynman}
    \vertex (i1) {$X_+$};
    \vertex at ($(i1) + (0.0cm, -1.5cm)$) (i2) {$X_-$};
    \vertex at ($(i1) + (0.0cm, -3cm)$) (i3) {$X_+$};
    \vertex at ($(i1) + (3cm, -0cm)$) (f1) {$X_+$};
    \vertex at ($(i1) + (3cm, -3cm)$) (f2) {$\tilde{X}_3$};
    \vertex at ($(i1) + (1.5cm, -0cm)$) (m1);
    \vertex at ($(i1) + (1.5cm, -1.5cm)$) (m2);
    \vertex at ($(i1) + (1.5cm, -3cm)$) (m3);
    \diagram {
      {(i1),(f1)} -- [photon] (m1) -- [photon, edge label = $\tilde{X}_3$] (m2) -- [photon, edge label = $X_+$] (m3) -- [photon] {(i3),(f2)},
      (m2) -- [photon] (i2)      
    };
\end{feynman}
\end{tikzpicture}
\quad
%8
\begin{tikzpicture}
\begin{feynman}
    \vertex (i1) {$X_-$};
    \vertex at ($(i1) + (0.0cm, -1.5cm)$) (i2) {$X_+$};
    \vertex at ($(i1) + (0.0cm, -3cm)$) (i3) {$X_+$};
    \vertex at ($(i1) + (3cm, -0cm)$) (f1) {$\tilde{X}_3$};
    \vertex at ($(i1) + (3cm, -2.25cm)$) (f2) {$X_+$};
    \vertex at ($(i1) + (1.5cm, -0.25cm)$) (m1);
    \vertex at ($(i1) + (1.5cm, -2.25cm)$) (m2);
    \diagram {
      {(i1),(f1)} -- [photon] (m1) -- [photon, edge label = $X_+$] (m2) -- [photon] {(i2),(i3),(f2)},      
    };
\end{feynman}
\end{tikzpicture}
\quad
%9
\begin{tikzpicture}
\begin{feynman}
    \vertex (i1) {$X_+$};
    \vertex at ($(i1) + (0.0cm, -1.5cm)$) (i2) {$X_+$};
    \vertex at ($(i1) + (0.0cm, -3cm)$) (i3) {$X_-$};
    \vertex at ($(i1) + (3cm, -0cm)$) (f1) {$\tilde{X}_3$};
    \vertex at ($(i1) + (3cm, -2.25cm)$) (f2) {$X_+$};
    \vertex at ($(i1) + (1.5cm, -0.25cm)$) (m1);
    \vertex at ($(i1) + (1.5cm, -2.25cm)$) (m2);
    \diagram {
      {(i1),(f1)} -- [photon] (m1) -- [photon, edge label = $X_+$] (m2) -- [photon] {(i2),(i3),(f2)},      
    };
\end{feynman}
\end{tikzpicture}

      \end{center}
  \caption{Feynman diagrams for $X_+ X_+ X_-\rightarrow X_+ {\tilde X}_3$.}
  \label{SIMPdiagrams}
\end{figure}

The effective $3\rightarrow 2$ annihilation cross section in eq.~(\ref{Boltz}) is given by the channel, $X_+ X_+ X_-\rightarrow X_+ {\tilde X}_3$, as shown in Fig.~\ref{SIMPdiagrams}, with the corresponding cross section given by
\bea
 \lab\sigma v^2\rab_{X_+X_+X_-\rightarrow X_+\tilde{X}_3} 
&=& \frac{\cos^2(\theta'_X) g_X^6}{161243136\pi m_X^5}\bigg(\frac{m_{\tilde{X}_3}}{m_X}\bigg)^{14} \bigg(1-\frac{m_{\tilde{X}_3}^2}{16m_X^2}\bigg)^{3/2} \bigg(1-\frac{m_{\tilde{X}_3}^2}{4m_X^2}\bigg)^{-1/2}  \nonumber \\
&& \times \bigg(1+\frac{m_{\tilde{X}_3}^2}{2m_X^2}\bigg)^{-2}\bigg( 3A_1 + 4\cos(2\theta'_X)A_2 + \cos(4\theta'_X)A_3 \bigg)
\eea
where
\bea
A_1&=& 17 + 823\bigg(\frac{m_X}{m_{\tilde{X}_3}}\bigg)^2 + 5380 \bigg(\frac{m_X}{m_{\tilde{X}_3}}\bigg)^4 + 306672 \bigg(\frac{m_X}{m_{\tilde{X}_3}}\bigg)^6 - 1964704 \bigg(\frac{m_X}{m_{\tilde{X}_3}}\bigg)^8 \nonumber  \\
&&+ 6233600 \bigg(\frac{m_X}{m_{\tilde{X}_3}}\bigg)^{10}  - 3080192 \bigg(\frac{m_X}{m_{\tilde{X}_3}}\bigg)^{12} + 13860864 \bigg(\frac{m_X}{m_{\tilde{X}_3}}\bigg)^{14},   \\
A_2&=& 5 + 219 \bigg(\frac{m_X}{m_{\tilde{X}_3}}\bigg)^{2} + 180 \bigg(\frac{m_X}{m_{\tilde{X}_3}}\bigg)^{4} + 69152 \bigg(\frac{m_X}{m_{\tilde{X}_3}}\bigg)^{6} - 1351488 \bigg(\frac{m_X}{m_{\tilde{X}_3}}\bigg)^{8}  \nonumber  \\
&& + 6657120 \bigg(\frac{m_X}{m_{\tilde{X}_3}}\bigg)^{10}-8880832 \bigg(\frac{m_X}{m_{\tilde{X}_3}}\bigg)^{12} + 9142272 \bigg(\frac{m_X}{m_{\tilde{X}_3}}\bigg)^{14},   \\
A_3&=& 1 + 39\bigg(\frac{m_X}{m_{\tilde{X}_3}}\bigg)^{2} -228\bigg(\frac{m_X}{m_{\tilde{X}_3}}\bigg)^{4} +12400\bigg(\frac{m_X}{m_{\tilde{X}_3}}\bigg)^{6} -399648\bigg(\frac{m_X}{m_{\tilde{X}_3}}\bigg)^{8}  \nonumber  \\
&&+ 3725184\bigg(\frac{m_X}{m_{\tilde{X}_3}}\bigg)^{10} -12369536\bigg(\frac{m_X}{m_{\tilde{X}_3}}\bigg)^{12} + 22413312\bigg(\frac{m_X}{m_{\tilde{X}_3}}\bigg)^{14}.
\eea
Then, assuming that dark matter annihilates mainly due to $3\rightarrow 2$ channels, we can determine the relic density \cite{fdm,simpG} as
\bea
\Omega_X h^2\simeq 0.12\, \bigg( \frac{g_*}{10.75}\bigg)^{-3/4} \bigg(\frac{x_f}{15} \bigg)^2 \left(\frac{m_X/\alpha_{\rm eff}}{45\,{\rm MeV}}\right)^{3/2}
\eea
where $\alpha_{\rm eff}$ is defined from $\langle \sigma v^2_{\rm rel}\rangle_{3\rightarrow 2}\equiv \alpha^3_{\rm eff}/m^5_X$.

\begin{figure}
  \begin{center}
    \includegraphics[height=0.55\textwidth]{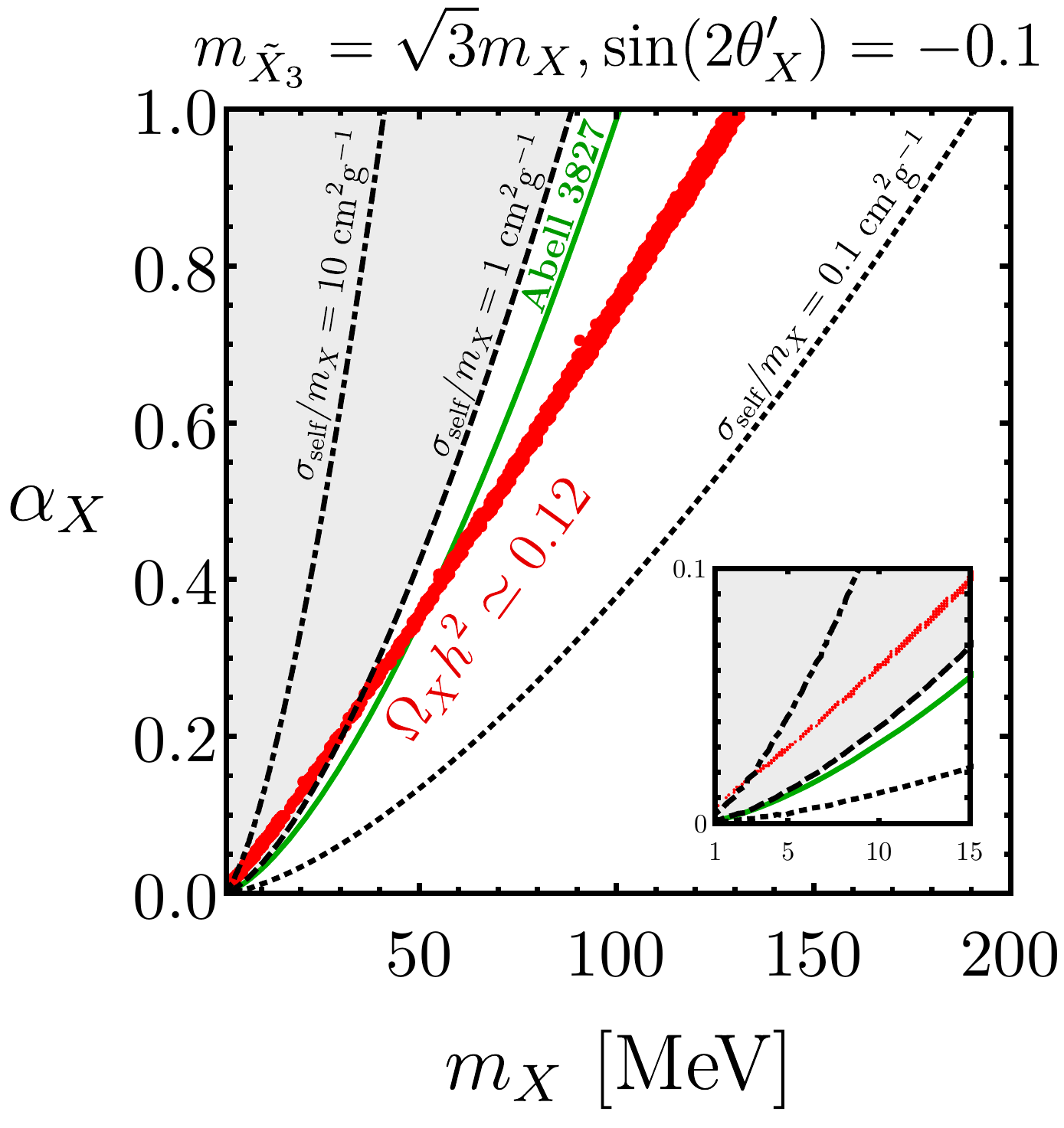}
  \end{center}
  \caption{Parameter space for $m_X$ vs $\alpha_X\equiv g^2_X/4\pi$ for the case with quadruplet dark Higgs. The correct relic density is satisfied in red region. Contours for DM self-scattering cross section with $\sigma_{\rm self}/m_X=0.1, 1, 10\,{\rm cm^2/g}$ are shown in dotted, dashed and dot-dashed lines, respectively.  We also indicated the favored parameter space in green line to explain the off-set of one of the galaxies in Abell 3827 \cite{abell3827,abell-kai,abell-recent} and the region disfavored by Bullet cluster in gray \cite{bullet}. }
  \label{simp1}
\end{figure}

In Fig.~\ref{simp1}, we showed the parameter space for $m_X$ and $\alpha_X\equiv g^2_X/(4\pi)$, satisfying the relic density condition in red region, dominantly due to SIMP channels. We took the VEV of quadruplet Higgs to get masses for non-abelian gauge bosons, for which $m_{{\tilde X}_3}\approx \sqrt{3}\, m_X$.  
Forbidden channels become dominant only for small $m_X$ or $\alpha_X$.  We also drew the contours with $\sigma_{\rm self}/m_X=0.1, 1, 10\,{\rm cm^2/g}$ in dotted, dashed and dot-dashed lines.  We note that the observed off-set of one of the galaxies of Abell 3827 \cite{abell3827,abell-kai} might indicate a large self-scattering cross section for dark matter, $\sigma_{\rm self}\cos\theta_i/m_X\sim 0.68\,{\rm cm^2/g}$ \cite{abell-recent}, where the cross section value depends on the unknown inclination angle $\theta_i$ of the galaxy's 3D motion with respect to the plane of the sky.
For comparison, taking $\theta_i=0$, we also showed the parameter space potentially favored by Abell 3827 in green line. The gray region is disfavored by Bullet cluster bound \cite{bullet}, $\sigma_{\rm self}/m_X\lesssim 1\,{\rm cm^2/g}$.

\begin{figure}
  \begin{center}
    \includegraphics[height=0.34\textwidth]{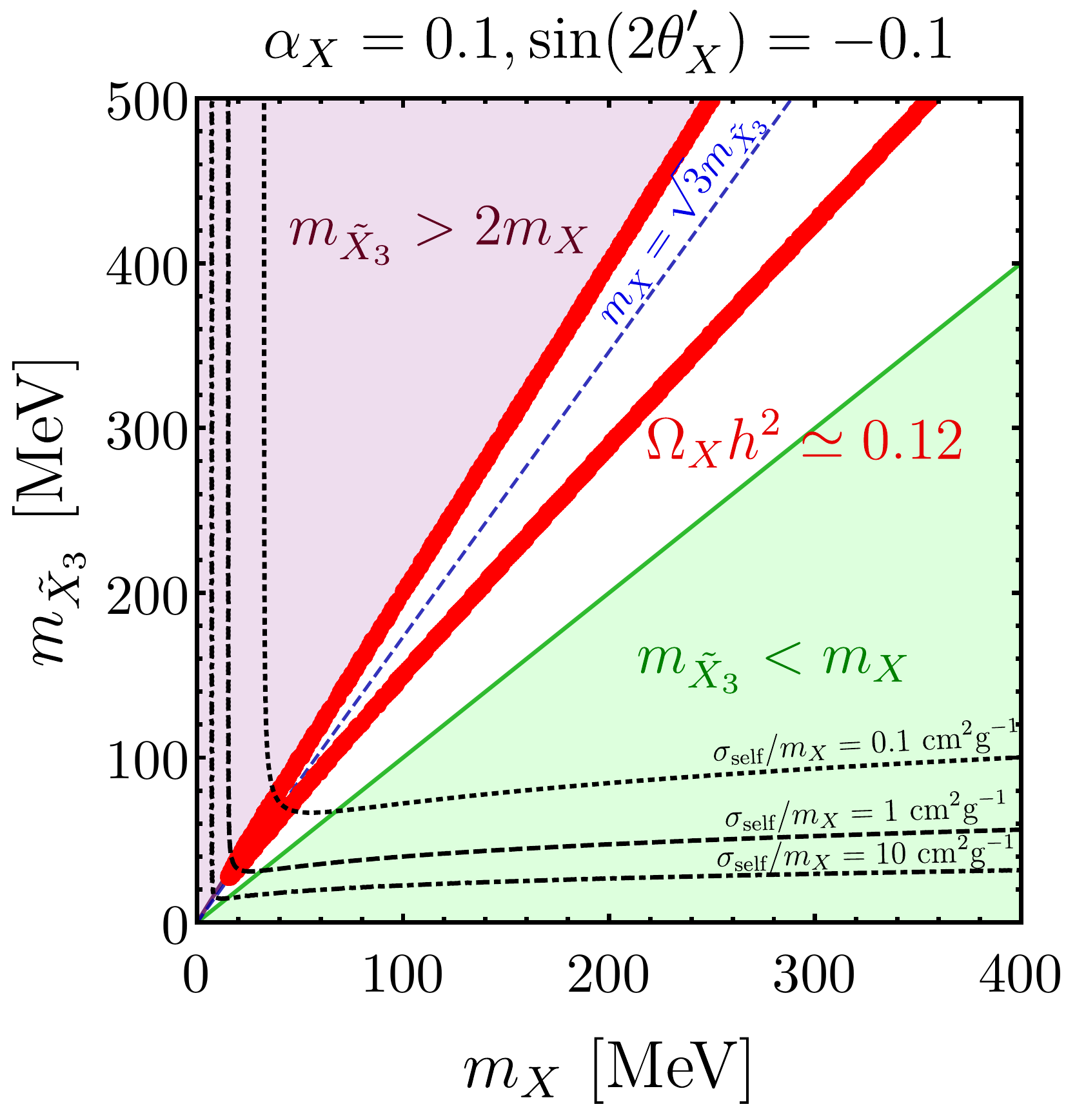} 
        \includegraphics[height=0.34\textwidth]{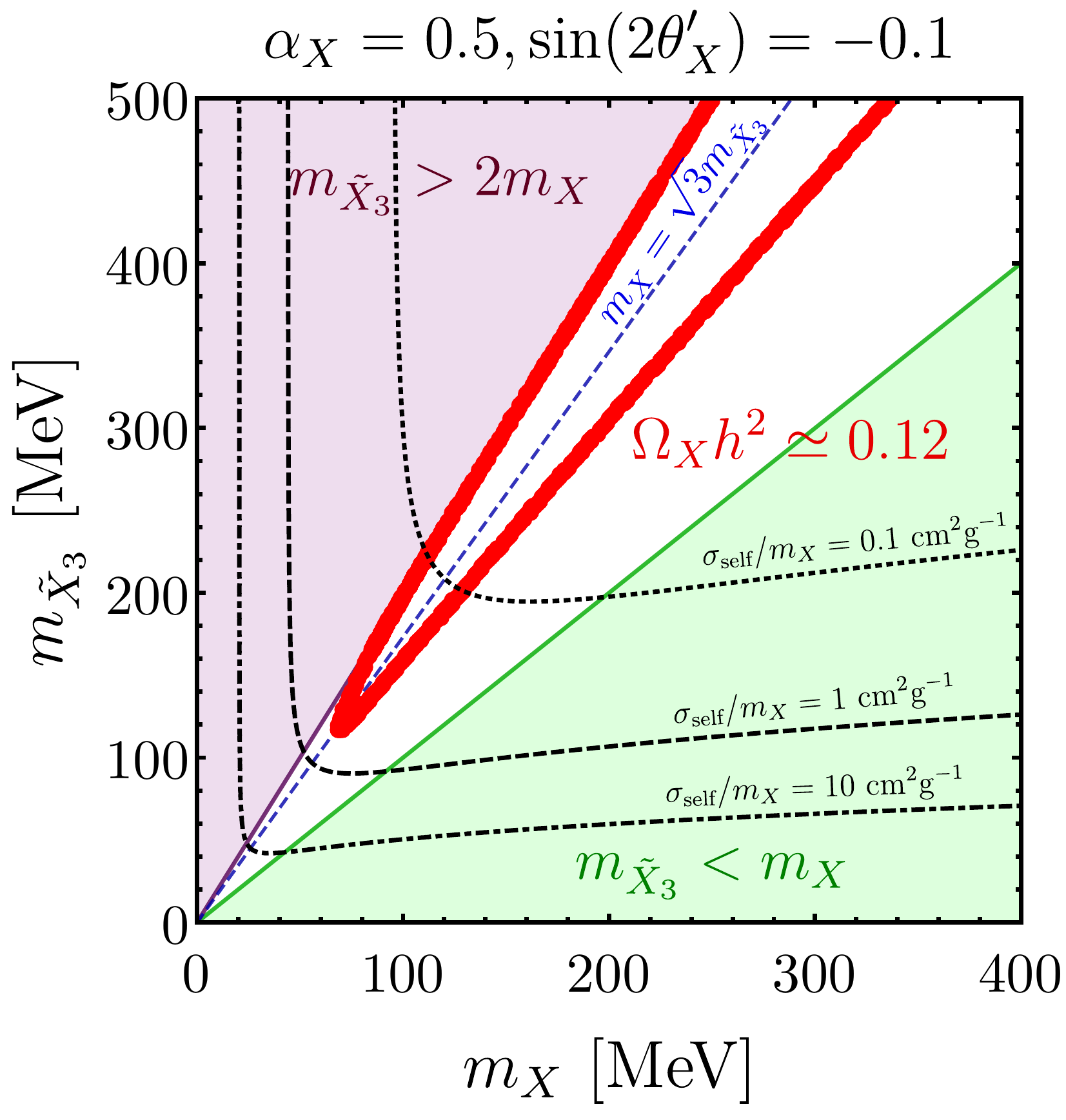} 
            \includegraphics[height=0.34\textwidth]{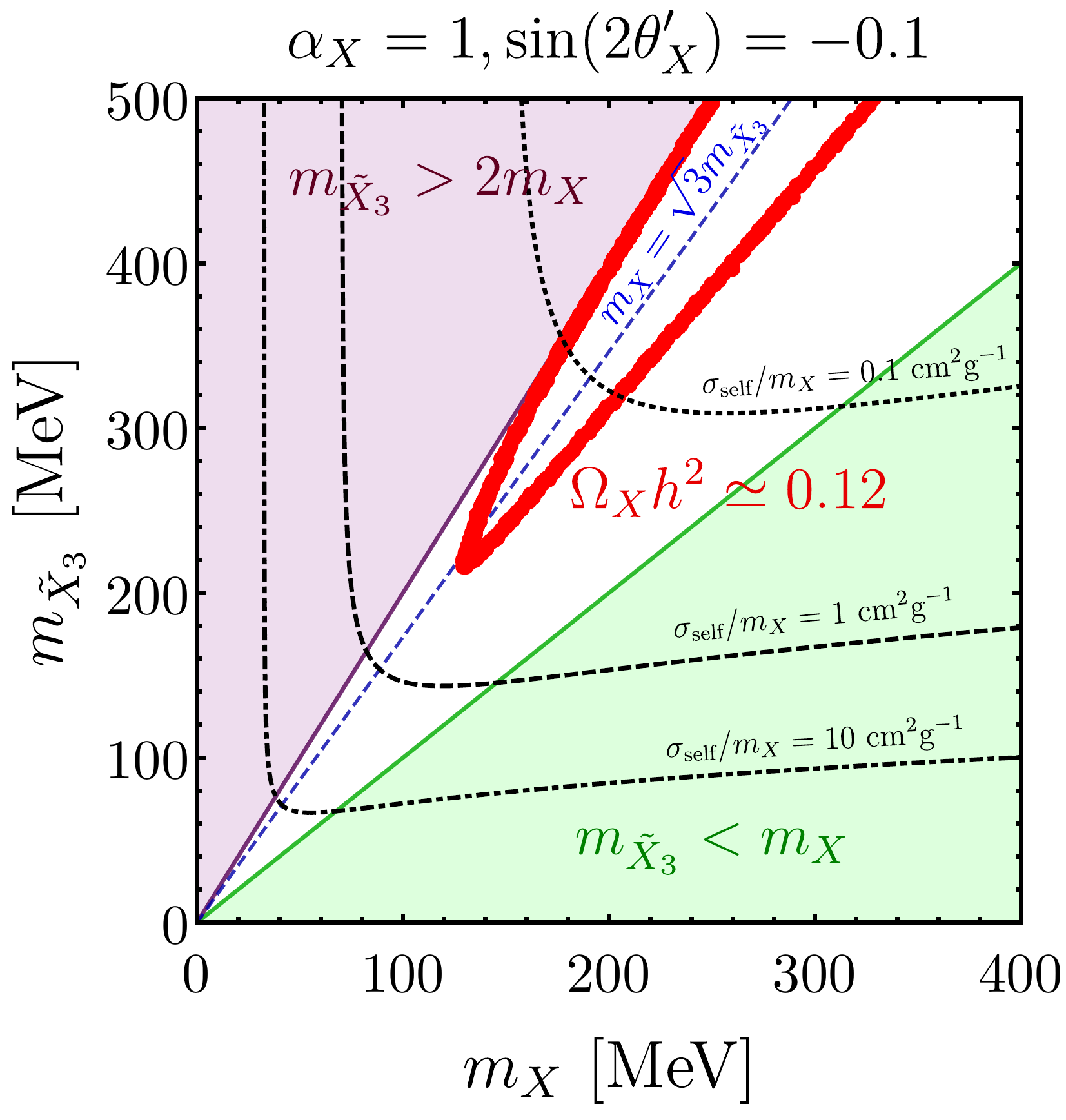}
  \end{center}
  \caption{Parameter space for $m_X$ vs $m_{{\tilde X}_3}$. Both SIMP and forbidden $2\rightarrow 2$ channels are included for determining the relic density. Contours with DM self-scattering cross section, $\sigma_{\rm self}/m_X=0.1, 1, 10\,{\rm cm^2/g}$ are shown in dotted, dashed and dotdashed lines, respectively. SIMP processes are not allowed kinematically in the upper purple region whereas $2\rightarrow 2$ annihilations in the dark sector are dominant in the lower green region. The blue dashed line corresponds to the mass relation, $m_{{\tilde X}_3}=\sqrt{3} \, m_X$, for the quadruplet dark Higgs.}
  \label{simp2}
\end{figure}

In Fig.~\ref{simp2}, we also depicted the parameter space for $m_X$ and $m_{{\tilde X}_3}$, satisfying the correct relic density in red regions. We took different values for $\alpha_X=0.1, 0.5, 1$, from left to right. We included both the SIMP and forbidden $2\rightarrow 2$ channels for determining the relic density. Forbidden channels are important for $m_{{\tilde X}_3}\lesssim 1.6m_X$, as will be discussed in the next subsection.  In each plot of Fig.~\ref{simp2}, the SIMP channels are kinematically closed in the upper purple region because $m_{{\tilde X}_3}>2m_X$, whereas the $2\rightarrow 2$ annihilations such as $X_+ X_-\rightarrow {\tilde X}_3{\tilde X}_3$ are open and dominant in the lower green region.  For comparison, the mass relation, $m_{{\tilde X}_3}\approx \sqrt{3} m_X$, is satisfied along the blue dashed line. Below the blue dashed line, the forbidden channels tend to contribute to the DM annihilations dominantly. But, above the blue dashed line, the SIMP channels are dominant, and the relic density is saturated close to $m_{{\tilde X}_3}\sim 2m_X$, due to the $t$-channel diagrams. This behavior is regularized by a relatively large velocity of dark matter during freeze-out.

\subsection{Forbidden $2\rightarrow 2$ annihilations}

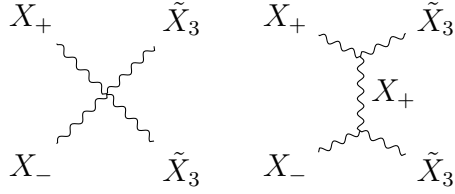
\begin{figure}
  \begin{center}
    \begin{tikzpicture}
\begin{feynman}
    \vertex (i1) {$X_+$};
    \vertex at ($(i1) + (0.0cm, -2cm)$) (i2) {$X_-$};
    \vertex at ($(i1) + (2cm, +0.0cm)$) (f1) {$\tilde{X}_3$};
    \vertex at ($(i1) + (2cm, -2cm)$) (f2) {$\tilde{X}_3$};
    \vertex at ($(i1) + (1cm, -1cm)$) (m1);
    \diagram {
      {(i1),(i2)} -- [photon] (m1) -- [photon] {(f1),(f2)},    
    };
\end{feynman}
\end{tikzpicture}
%2
\quad
\begin{tikzpicture}
\begin{feynman}
    \vertex (i1) {$X_+$};
    \vertex at ($(i1) + (0.0cm, -2cm)$) (i2) {$X_-$};
    \vertex at ($(i1) + (2cm, +0.0cm)$) (f1) {$\tilde{X}_3$};
    \vertex at ($(i1) + (2cm, -2cm)$) (f2) {$\tilde{X}_3$};
    \vertex at ($(i1) + (1cm, -0.5cm)$) (m1);
    \vertex at ($(i1) + (1cm, -1.5cm)$) (m2);
    \diagram {
      {(i1),(f1)} -- [photon] (m1) -- [photon, edge label = $X_+$] (m2) -- [photon] {(i2),(f2)},      
    };
\end{feynman}
\end{tikzpicture}
  \end{center}
  \caption{Feynman diagrams for $X_+ X_- \rightarrow \tilde{X}_3 \tilde{X}_3$. }
  \label{forbdiagrams}
\end{figure}

Next, the forbidden $2\rightarrow 2$ annihilation channels are shown in Fig.~\ref{forbdiagrams}.
For the inverse process of $2\rightarrow 2$ annihilation, i.e. $\tilde{X}_3 \tilde{X}_3 \rightarrow X_+ X_-$, in the dark sector, the corresponding cross section is given by
\bea
 \langle\sigma v_{\rm rel}\rangle_{\tilde{X}_3 \tilde{X}_3 \rightarrow X_+ X_-} &=&  
\frac{\cos^4(\theta_X')g_X^4}{144\pi m_X^2}\bigg(\frac{m_{\tilde{X}_3}}{m_X}\bigg)^6\sqrt{1-\frac{m_X^2}{m_{\tilde{X}_3^2}}} \nonumber  \\
&& \quad \times\bigg[ 1 - 2\bigg( \frac{m_X}{m_{\tilde{X}_3}} \bigg)^2+\bigg( \frac{m_X}{m_{\tilde{X}_3}} \bigg)^4+64\bigg( \frac{m_X}{m_{\tilde{X}_3}} \bigg)^6-71\bigg( \frac{m_X}{m_{\tilde{X}_3}} \bigg)^8 \nonumber \\
&&\qquad\quad +88\bigg( \frac{m_X}{m_{\tilde{X}_3}} \bigg)^{10} +48\bigg( \frac{m_X}{m_{\tilde{X}_3}} \bigg)^{12} \bigg]. 
\eea

For dominance with forbidden channels in  the Boltzmann equation (\ref{Boltz}), the solution for the DM abundance \cite{fdm} becomes
\bea
Y_X(\infty)= \frac{x_f}{\lambda}\, e^{2\Delta x_f}\, f(\Delta,x_f)
\eea
where $\lambda\equiv s(m_X)/H(m_X)$ with $s(m_X)$ and $H(m_X)$ being entropy density and  Hubble parameter, respectively, evaluated at $T=m_X$, and
\bea
f(\Delta,x_f)= \bigg[ \frac{1}{2}\langle \sigma v_{\rm rel} \rangle_{\tilde{X}_3 \tilde{X}_3 \rightarrow X_+ X_-}  (1+\Delta)^3\bigg( 1-2\Delta \, x_f \, e^{2\Delta x_f}\int_{2\Delta x_f}^\infty dt\, t^{-1} e^{-t} \bigg) \bigg]^{-1}
\eea
and  $\Delta\equiv  (m_{\tilde{X}_3} - m_X)/m_X$.
Then, forbidden channels can determine the relic abundance \cite{fdm}  as follows,
\bea
\Omega_X h^2\simeq  0.12\,  \bigg(\frac{g_*}{10.75}\bigg)^{-1/2} \bigg( \frac{x_f}{15}\bigg) \, e^{2(\Delta-0.6) x_f}\, \bigg(\frac{m_X}{100\,{\rm MeV}}\bigg)^2 \left(\frac{4.6\times 10^{-2}}{\beta_{\rm eff}}\right)^2
\eea
where $\beta_{\rm eff}$ is defined from $ f(\Delta,x_f)\equiv m^2_X/\beta^2_{\rm eff}$ and we took $x_f\simeq 15$ in the exponent. 
As a consequence, as far as $\Delta\lesssim 0.6$, the forbidden channels can be efficient enough even for a small self-interaction to produce a correct relic density.  This is the case with the triplet dark Higgs, for which $\Delta\approx \sqrt{2}-1$.
In cases with quadruplet and quintuplet dark Higgs fields, $\Delta$ is larger so that SIMP channels are dominant in determining the relic density.

\begin{figure}
  \begin{center}
    \includegraphics[height=0.26\textwidth]{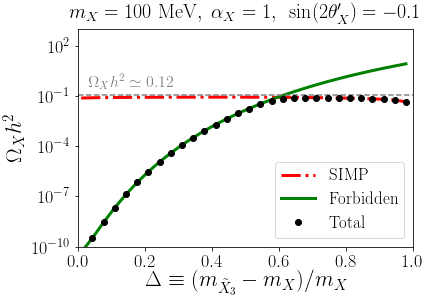} 
        \includegraphics[height=0.26\textwidth]{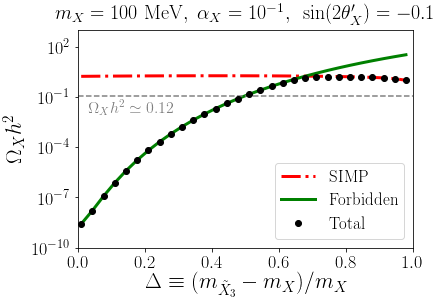}.   \vspace{0.5cm} \\
            \includegraphics[height=0.26\textwidth]{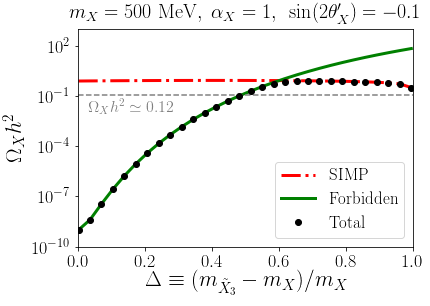}
                \includegraphics[height=0.26\textwidth]{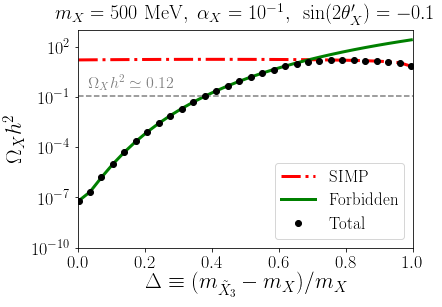}
  \end{center}
  \caption{Relic density as a function of $\Delta=(m_{{\tilde X}_3}-m_X)/m_X$. SIMP channels only are in red and forbidden channels only are in green, while both channels are included in black. }
  \label{forbidden1}
\end{figure}

In Fig.~\ref{forbidden1}, we showed the relic density as a function of the mass difference, $\Delta=(m_{{\tilde X}_3}-m_X)/m_X$ in black dotted line. In each plot, SIMP channels only are assumed in red line and forbidden channels only are assumed in green line. We have included both SIMP and forbidden channels in black line.  In all the cases for $m_X\sim 100-500\,{\rm MeV}$, SIMP channels become dominant for $\Delta\gtrsim 0.6$ or $m_{{\tilde X}_3}\gtrsim 1.6m_X$. Therefore, from eq.~(\ref{split}), $\Delta\simeq \sqrt{2I}-1$, so we need that dark Higgs fields with $I=\frac{3}{2}$ or $2$ determine the $SU(2)_X$ gauge boson masses dominantly. 
Otherwise, the relic density is determined mainly by forbidden channels.

\begin{figure}
  \begin{center}
            \includegraphics[height=0.26\textwidth]{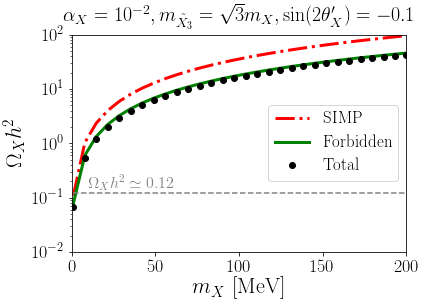}
               \includegraphics[height=0.26\textwidth]{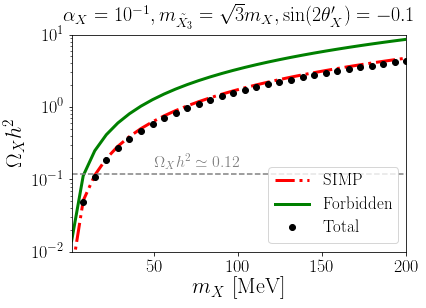}   \vspace{0.5cm}\\
                \includegraphics[height=0.26\textwidth]{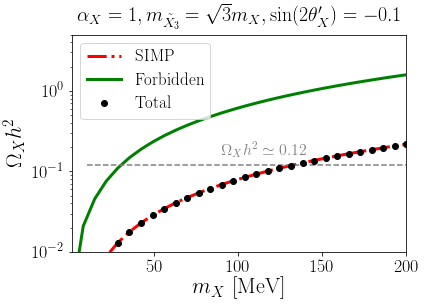}
                    \includegraphics[height=0.26\textwidth]{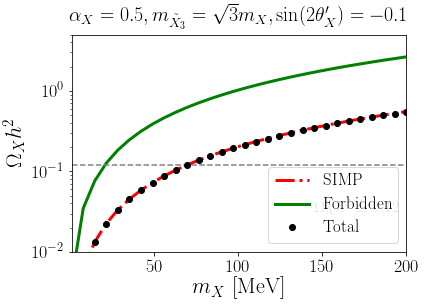} 
  \end{center}
  \caption{Relic density as a function of $m_X$ for the case with quadruplet dark Higgs. SIMP channels only are in red and forbidden channels only are in green, while both channels are included in black.}
  \label{forbidden2}
\end{figure}

In Fig.~\ref{forbidden2}, we drew similar plots as in Fig.~\ref{forbidden1}, but now for the relic density as a function of $m_X$ for the fixed mass relation, $m_{{\tilde X}_3}= \sqrt{3} \, m_X$, for the nonzero VEV of quadruplet Higgs. In this case, as we increase $\alpha_X$ from $0.01
$ to $0.1, 0.5$ and $1$, clockwise, showing the dominance of SIMP channels in determining the relic density for $\alpha_X\gtrsim 0.1$. 
For a correct relic density, we need a sizable $\alpha_X$ as in the plots in the lower panel, in which case the SIMP channels are dominant.

\subsection{DM self-scattering}

\begin{figure}
  \begin{center}
          \begin{tikzpicture}
\begin{feynman}
    \vertex (i1) {$X_+$};
    \vertex at ($(i1) + (0.0cm, -2cm)$) (i2) {$X_+$};
    \vertex at ($(i1) + (2cm, +0.0cm)$) (f1) {$X_+$};
    \vertex at ($(i1) + (2cm, -2cm)$) (f2) {$X_+$};
    \vertex at ($(i1) + (1cm, -1cm)$) (m1);
    \diagram {
      {(i1),(i2)} -- [photon] (m1) -- [photon] {(f1),(f2)},    
    };
\end{feynman}
\end{tikzpicture}
%2
\begin{tikzpicture}
\begin{feynman}
    \vertex (i1) {$X_+$};
    \vertex at ($(i1) + (0.0cm, -2cm)$) (i2) {$X_+$};
    \vertex at ($(i1) + (2cm, +0.0cm)$) (f1) {$X_+$};
    \vertex at ($(i1) + (2cm, -2cm)$) (f2) {$X_+$};
    \vertex at ($(i1) + (1cm, -0.5cm)$) (m1);
    \vertex at ($(i1) + (1cm, -1.5cm)$) (m2);
    \diagram {
      {(i1),(f1)} -- [photon] (m1) -- [photon, edge label = $\tilde{X}_3$] (m2) -- [photon] {(i2),(f2)},      
    };
\end{feynman}
\end{tikzpicture}
%3
\begin{tikzpicture}
\begin{feynman}
    \vertex (i1) {$X_+$};
    \vertex at ($(i1) + (0.0cm, -2cm)$) (i2) {$X_-$};
    \vertex at ($(i1) + (2cm, +0.0cm)$) (f1) {$X_+$};
    \vertex at ($(i1) + (2cm, -2cm)$) (f2) {$X_-$};
    \vertex at ($(i1) + (1cm, -1cm)$) (m1);
    \diagram {
      {(i1),(i2)} -- [photon] (m1) -- [photon] {(f1),(f2)},   
    };
\end{feynman}
\end{tikzpicture}
%4
\begin{tikzpicture}
\begin{feynman}
    \vertex (i1) {$X_+$};
    \vertex at ($(i1) + (0.0cm, -2cm)$) (i2) {$X_-$};
    \vertex at ($(i1) + (2cm, +0.0cm)$) (f1) {$X_+$};
    \vertex at ($(i1) + (2cm, -2cm)$) (f2) {$X_-$};
    \vertex at ($(i1) + (1cm, -0.5cm)$) (m1);
    \vertex at ($(i1) + (1cm, -1.5cm)$) (m2);
    \diagram {
      {(i1),(f1)} -- [photon] (m1) -- [photon, edge label = $\tilde{X}_3$] (m2) -- [photon] {(i2),(f2)},     
    };
\end{feynman}
\end{tikzpicture}
%5
\begin{tikzpicture}
\begin{feynman}
    \vertex (i1) {$X_+$};
    \vertex at ($(i1) + (0.0cm, -2cm)$) (i2) {$X_-$};
    \vertex at ($(i1) + (2cm, +0.0cm)$) (f1) {$X_+$};
    \vertex at ($(i1) + (2cm, -2cm)$) (f2) {$X_-$};
    \vertex at ($(i1) + (0.5cm, -1cm)$) (m1);
    \vertex at ($(i1) + (1.5cm, -1cm)$) (m2);
    \diagram {
      {(i1),(i2)} -- [photon] (m1) -- [photon, edge label = $\tilde{X}_3$] (m2) -- [photon] {(f1),(f2)},      
    };
\end{feynman}
\end{tikzpicture}
  \end{center}
  \caption{Feynman diagrams for DM self-scattering channels, $X_\pm X_\pm\rightarrow X_\pm X_\pm $ and $X_+ X_-\rightarrow X_+ X_-$.}
  \label{selfscatt}
\end{figure}
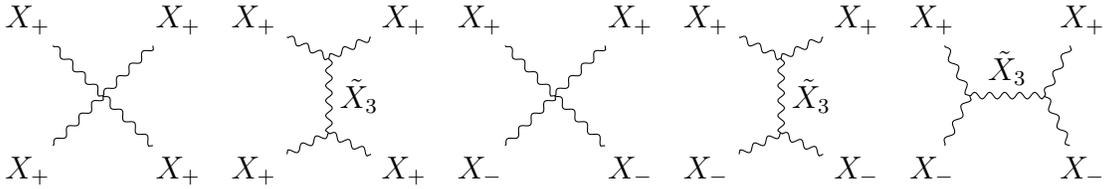

The DM $2\rightarrow 2$ self-scattering channels are given in Fig.~\ref{selfscatt}.
Then, the effective DM self-scattering cross section is given by
\bea
\sigma_{\rm self}&=& \frac{1}{4}\bigg(\sigma_{X_+X_-\rightarrow X_+X_-} + \sigma_{X_+X_+\rightarrow X_+X_+} + \sigma_{X_-X_-\rightarrow X_-X_-}\bigg) \nonumber \\
&=& \frac{g_X^4}{96\pi m_X^2}\bigg[ 3 + 2\bigg(\frac{m_X}{m_{\tilde{X}_3}}\cos(\theta_X')\bigg)^2 + 22\bigg(\frac{m_X}{m_{\tilde{X}_3}}\cos(\theta_X')\bigg)^4 \bigg]
\eea
with
\bea
\sigma_{X_+X_+\rightarrow X_+X_+}& =& \sigma_{X_-X_-\rightarrow X_-X_-} \nonumber \\
&=& \frac{g_X^4}{96\pi m_X^2}\bigg[ 3 +8\bigg(\frac{m_X}{m_{\tilde{X}_3}}\cos(\theta_X')\bigg)^2 +32\bigg(\frac{m_X}{m_{\tilde{X}_3}}\cos(\theta_X')\bigg)^4 \bigg], \\
\sigma_{X_+X_-\rightarrow X_+X_-} &=& \frac{g_X^4}{48\pi m_X^2}\bigg[  3-4\bigg(\frac{m_X}{m_{\tilde{X}_3}}\cos(\theta_X')\bigg)^2 + 12\bigg(\frac{m_X}{m_{\tilde{X}_3}}\cos(\theta_X')\bigg)^4 \bigg].
\eea
Then, the DM self-scattering with $\sigma_{\rm self}/m_X=0.1-10\,{\rm cm^2/g}$ can explain small-scale problems at galaxies such as core-cusp problem, too-big-to-fail problem, etc. On the other hand, we also note that the DM self-scattering cross section is constrained to be $\sigma_{\rm self}/m_X\lesssim 1\,{\rm cm^2/g}$ by either Bullet cluster \cite{bullet} or halo shapes/ellipticity \cite{halo}.

\section{$Z'$ portal couplings for dark matter}

Dark matter, if its abundance is determined mainly by SIMP processes, must be in kinetic equilibrium with another species in the dark or visible sectors for the structure formation. 
In order not to introduce the temperature of dark matter as an independent parameter, we consider the case that dark matter is equilibriated by the elastic scattering between dark matter and light particles in the SM.

In this section, we discuss the conditions for kinetic equilibrium in the presence of a gauge kinetic mixing between $Z'$ and SM hypercharge gauge bosons, and consider various constraints on the model, coming from the consistency of SIMP scenarios to experimental bounds such as direct detection and collider searches.

\subsection{General current interactions with $Z'$ portal}

In the presence of a gauge kinetic mixing, we need to consider the full basis of neutral gauge bosons, including neutral dark gauge bosons and those in the SM, i.e. $B_\mu$ and $W_{3\mu}$.
For a zero dark Weinberg angle, we can easily  diagonalize the gauge kinetic terms and mass terms for neutral gauge bosons only by the $4\times 4$ rotation matrix, $O_W$, as in usual $Z'$ portal models.  Even with a nonzero dark Weinberg angle, we can still diagonalize the mass matrix by approximate rotations with  $O_W$, followed by the $2\times 2$ rotation matrix, $O_X$,  in the limit of a small gauge kinetic mixing. Then, the gauge bosons in the interaction basis, $(B_\mu, W_{3\mu},Z'_\mu,X_{3\mu})$, can be written in terms of approximate mass eigenstates $({\tilde A}_\mu,{\tilde Z}_\mu, {\tilde Z}'_\mu,{\tilde X}_{3\mu})$, as follows,
\begin{equation}
\left(
\begin{array}{c}
 B_\mu \\
 W_{3\mu} \\
 Z'_\mu \\
 X_{3\mu} \\
\end{array}
\right)=O_W O_X \left(
\begin{array}{c}
 \tilde{A}_\mu \\
 \tilde{Z}_\mu \\
 \tilde{Z}'_\mu \\
 \tilde{X}_{3\mu} \\
\end{array}
\right) \label{gaugerot}
\end{equation}
with
\bea
O_W&=&\left(
\begin{array}{cccc}
 c_W & t_{\xi } s_{\zeta }-s_W c_{\zeta } & -s_W s_{\zeta }-t_{\xi } c_{\zeta } & 0 \\
 s_W & c_W c_{\zeta } & c_W s_{\zeta } & 0 \\
 0 & -s_{\zeta }/c_{\xi } & c_{\zeta }/c_{\xi } & 0 \\
 0 & 0 & 0 & 1 \\
\end{array}
\right), \\
O_X&=& \left(
\begin{array}{cccc}
 1 & 0 & 0 & 0 \\
 0 & 1 & 0 & 0 \\
 0 & 0 & c_{\theta _X'} & -s_{\theta _X'} \\
 0 & 0 & s_{\theta _X'} & c_{\theta _X'} \\
\end{array}
\right).
\eea
Here, $c_\xi\equiv \cos\xi$, $t_\xi\equiv \tan\xi$,  $c_W\equiv \cos\theta_W$,  $s_W\equiv \sin\theta_W$, and $\zeta$ is the approximate mixing angle between $Z'$ and $Z$ bosons, given \cite{z3model} by
\bea
\tan(2\zeta)=\frac{m^2_Z s_W \sin(2\xi)}{m^2_{Z'}-m^2_Z (c^2_{\xi}-s^2_W s^2_\xi)}.
\eea
In the limit of $m^2_{Z'}\ll m^2_Z$ and $|\xi|\ll 1$, we obtain $\zeta\approx -s_W \xi$. 
Moreover, the approximate mass eigenvalues for $Z$-like and $Z'$-like gauge bosons are given \cite{z3model} by
\bea
m^2_{1,2}=\frac{1}{2} \left[ m^2_Z (1+s^2_W t^2_\xi)+ m^2_{Z'}/c^2_\xi\pm \sqrt{\Big(m^2_Z(1+s^2_W t^2_\xi)+m^2_{Z'}/c^2_\xi\Big)^2-4m^2_Z m^2_{Z'}/c^2_\xi}\right].
\eea

The current interactions in the interaction basis are given by
\bea
{\cal L}_{\rm EM/NC}&=&g_X X_{3\mu} J^\mu_{X_3} + g_{Z'} Z'_\mu J^\mu_{Z'} \nonumber \\ 
&&+e(s_W W_{3\mu}+c_W B_\mu) J^\mu_{\rm EM} +\frac{e}{2s_W c_W}\,(c_W W_{3\mu}- s_W B_\mu )J^\mu_Z
\eea
where $J_{X_3}^\mu$ and $J_{Z^\prime}^\mu$ are dark-neutral and $Z'$ currents, and $ J^\mu_{\rm EM} $ and $J^\mu_Z$ are electromagnetic and neutral currents. 
Then, using eq.~(\ref{gaugerot}), we can rewrite the above current interactions in the basis of mass eigenstates, at first order in $\varepsilon\equiv c_W t_\xi\approx c_W\xi$, as
 \bea
{\cal L}_{\rm EM/NC}&=& e \tilde{A}_\mu J^\mu_{\rm EM} + \tilde{Z}_\mu  \bigg[ \frac{e}{2s_Wc_W} J^\mu_Z + \varepsilon g_{Z^\prime} t_W  J^\mu_{Z^\prime} \bigg] \nonumber \\
&& +\tilde{Z}^\prime_\mu \bigg[ g_X \sin (\theta_X') J_{X_3}^\mu +g_{Z^\prime} \cos (\theta_X') J_{Z^\prime}^\mu -e \varepsilon \cos (\theta_X') J_\text{EM}^\mu   \bigg] \\
&& +\tilde{X}_{3 \mu} \bigg[ g_X \cos (\theta_X') J_{X_3}^\mu -g_{Z^\prime} \sin (\theta_X') J_{Z^\prime}^\mu +e \varepsilon \sin (\theta_X') J_\text{EM}^\mu   \bigg]. \nonumber
\label{eq:IntCurrentsNVDM}
\eea
Therefore, we find that dark-neutral gauge bosons, ${\tilde Z}'$ and ${\tilde X}_3$, couple to electromagnetic currents and dark-neutral currents as well as $Z'$ currents. 
Since ${\tilde X}_3$ has mass comparable to dark matter mass by the approximate custodial symmetry in the dark sector, it can play an important role for the kinetic equilibrium of dark matter by elastic scattering.

\begin{figure}
  \begin{center}
            \includegraphics[height=0.35\textwidth]{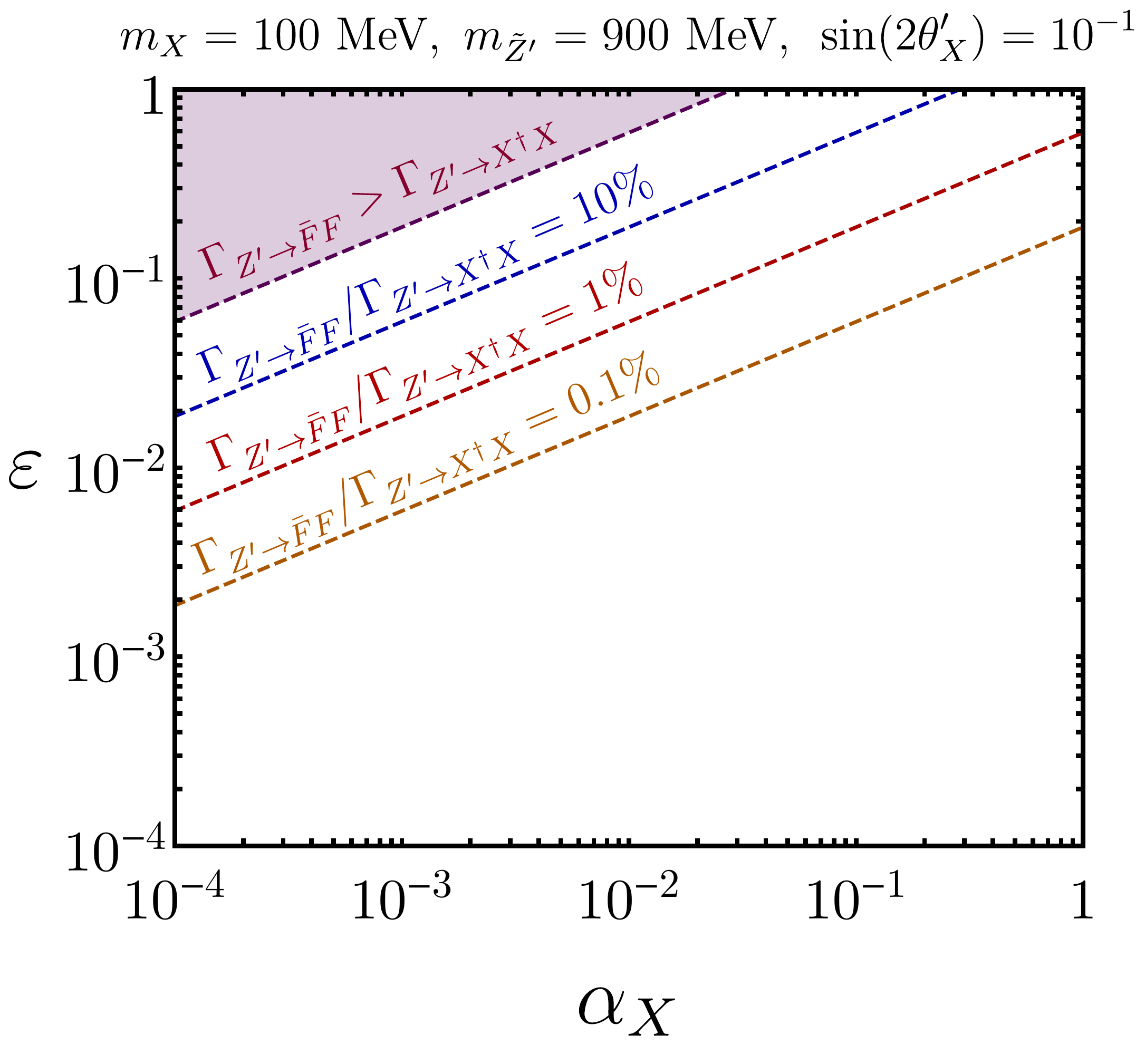}
               \includegraphics[height=0.35\textwidth]{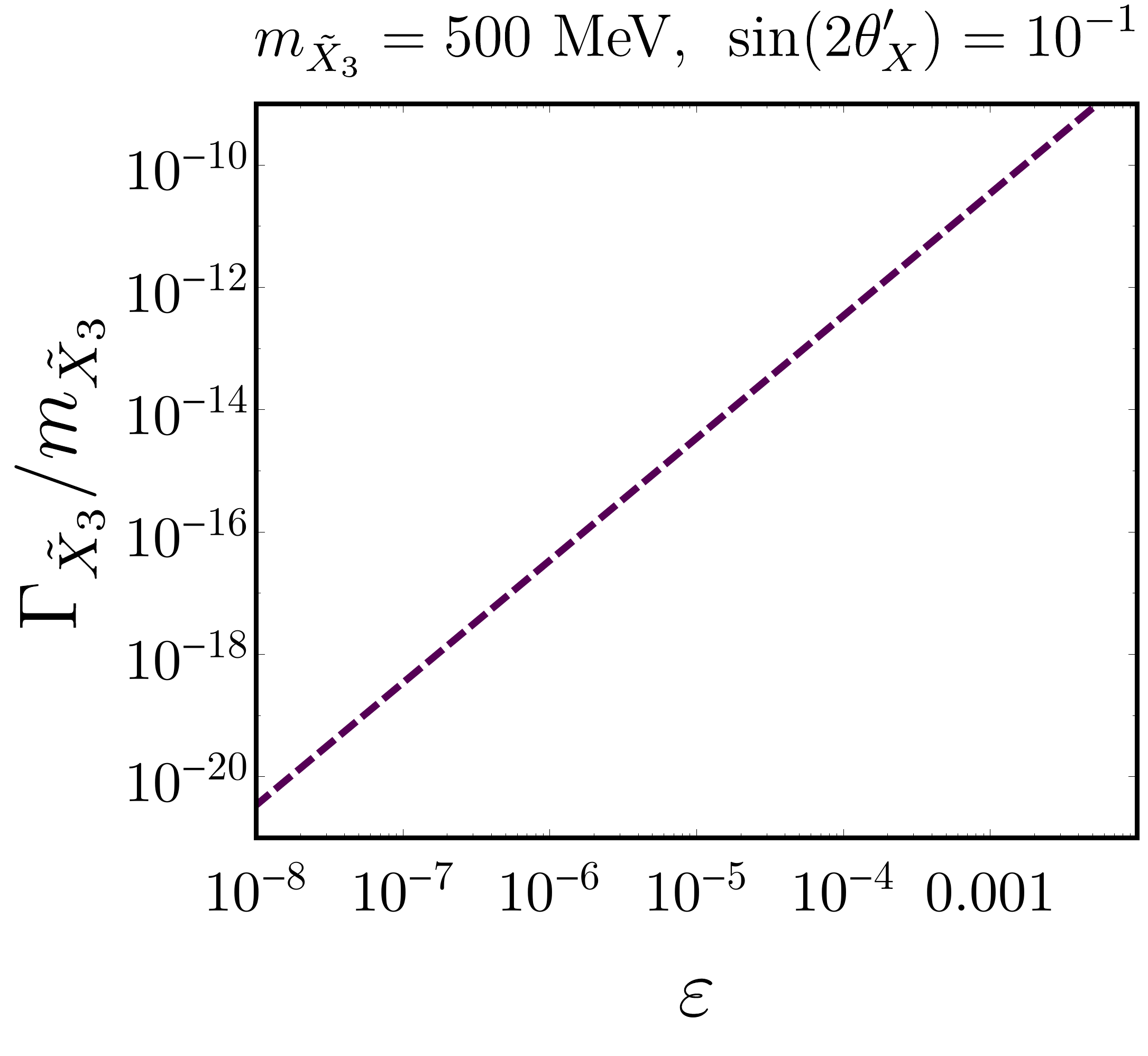}  
  \end{center}
  \caption{(Left) Ratio of visible to invisible decay rates of ${\tilde Z}'$ in the parameter space for $\varepsilon$ and $\alpha_X$. (Right) Ratio of decay rate to mass of ${\tilde X}_3$ as a function of $\varepsilon$.  }
  \label{decays}
\end{figure}

As a result, $\tilde{Z}'$ can decay into either a lepton pair or a pair of dark matter particles.
Then, the total decay width of $\tilde{Z}'$ is given by 
\bea
\Gamma_{\tilde{Z}'}=\Gamma_{\tilde{Z}'\rightarrow f{\bar f}}+\Gamma_{\tilde{Z}'\rightarrow X^+X^-}
\eea
with
\bea
\Gamma_{\tilde{Z}'\rightarrow f{\bar f}} &=& \frac{N_c Q_f^2 \epsilon^2 \cos^2(\theta_X') (2m_{f}^2+m_{\tilde{Z}'}^2)}{12\pi m_{\tilde{Z}'}}\sqrt{1-\frac{4m_{f}^2}{m_{\tilde{Z}'}^2}}, \\
\Gamma_{\tilde{Z}'\rightarrow X^+X^-} &=& \frac{g_X^2 \sin^2(\theta_X')m_{\tilde{Z}'}}{192\pi}\bigg(\frac{m_{\tilde{Z}'}}{m_X}\bigg)^4\sqrt{1-\frac{4m_X^2}{m_{\tilde{Z}'}^2}}\, \bigg[ 1+16\bigg(\frac{m_X}{m_{\tilde{Z}'}}\bigg)^2 \nonumber \\
&&-68\bigg(\frac{m_X}{m_{\tilde{Z}'}}\bigg)^4-48\bigg(\frac{m_X}{m_{\tilde{Z}'}}\bigg)^6 \bigg].
\eea
On the other hand, ${\tilde X}_3$ has mass $m_{{\tilde X}_3}<2m_X$ for SIMP processes to be kinematically open. So, ${\tilde X}_3$ decays dominantly into a pair of leptons with $f$ being electron or muon, with the decay rate given by 
\be
\Gamma_{\tilde{X}_3\rightarrow f{\bar f}} = \frac{e^2\epsilon^2\sin^2(\theta_X')(2m_f^2+m_{\tilde{X}_3}^2)}{12\pi m_{\tilde{X}_3}}\sqrt{1-\frac{4m_f^2}{m_{\tilde{X}_3}^2}}.
\ee 
For light dark-neutral gauge bosons of mass around QCD scale, in particular, ${\tilde X}_3$, whose mass is close to DM mass by custodial symmetry, we also need to include the decay modes into hadronic states. The hadronic width is given by
\bea
\Gamma_{\tilde{X}_3\rightarrow {\rm hadrons}}= \frac{e^2\epsilon^2\sin^2(\theta_X')m_{{\tilde X}_3} }{12\pi}\, \cdot R(m^2_{{\tilde X}_3})
\eea
where $R(s)$ is the ratio of hadronic cross section to muonic cross section at tree level \cite{pdg} in $e^+e^-$ annihilations at center of mass energy $\sqrt{s}$,
\bea
R(s)\equiv \frac{\sigma(e^+e^-\rightarrow {\rm hadron})}{\sigma(e^+e^-\rightarrow \mu^+\mu^-)}.
\eea
When ${\tilde Z}'$ is light, we need to include the hadronic decays for ${\tilde Z}'$ instead of quark decays by
\bea
\Gamma_{\tilde{Z}'\rightarrow {\rm hadrons}}= \frac{e^2\epsilon^2\cos^2(\theta_X')m_{{\tilde Z}'} }{12\pi}\,\cdot R(m^2_{{\tilde Z}'}).
\eea
The ratio $R(s)$ can be significantly greater than unity for $600\,{\rm MeV}\lesssim m_{{\tilde X}_3}, m_{{\tilde Z}'}\lesssim 2\,{\rm GeV}$ near hadronic resonances such as $\rho,\omega,\phi, \rho'$, etc \cite{pdg}.  Apart from those resonance regions, we can still apply the limits from visible modes with dileptons in our later discussion. But, we note that care should be taken of in interpreting our results for visible modes of light dark-neutral gauge bosons near hadronic resonances. On the other hand, the limits from invisible decays in the later section will be robust in most of parameter space of our interest, because they will dominate once open.

On left of Fig.~\ref{decays}, we depicted contours of the ratio of visible to invisible decay rates of ${\tilde Z}'$ gauge boson in the parameter space for $\varepsilon$ vs $\alpha_X\equiv g^2_X/(4\pi)$. Here, we have fixed $m_X=100\,{\rm MeV}$, $m_{{\tilde Z}'}=1\,{\rm GeV}$ and $\sin(2\theta'_X)=0.1$. So, in most of the parameter space for $\alpha_X\gtrsim 0.1$ and $\varepsilon\lesssim 10^{-2}$, the visible decay of  ${\tilde Z}'$ is negligibly small. 
On right of Fig.~\ref{decays}, we also showed the ratio of decay rate to mass of ${\tilde X}_3$ as a function of $\varepsilon$, for $m_{{\tilde X}_3}=500\,{\rm MeV}$ and $\sin(2\theta'_X)=0.1$. In this case, for $m_X<m_{{\tilde X}_3}<2m_X$, $m_{{\tilde X}_3}$ decays visibly into a lepton pair, so the decay rate is doubly suppressed by the dark Weinberg angle and the gauge kinetic parameter.

\subsection{Kinetic equilibrium}

\begin{figure}
  \begin{center}
            \includegraphics[height=0.34\textwidth]{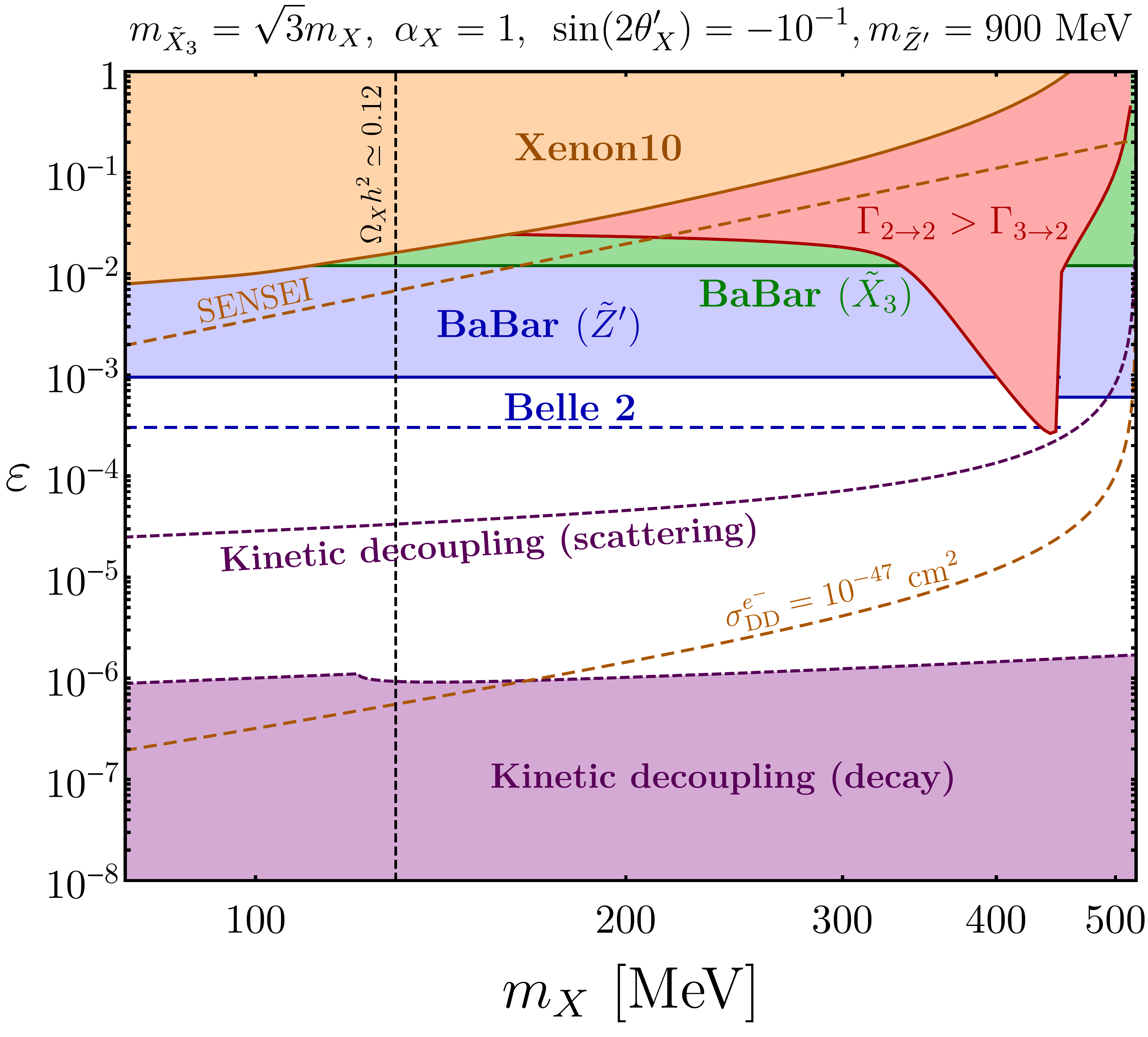} \quad
               \includegraphics[height=0.34\textwidth]{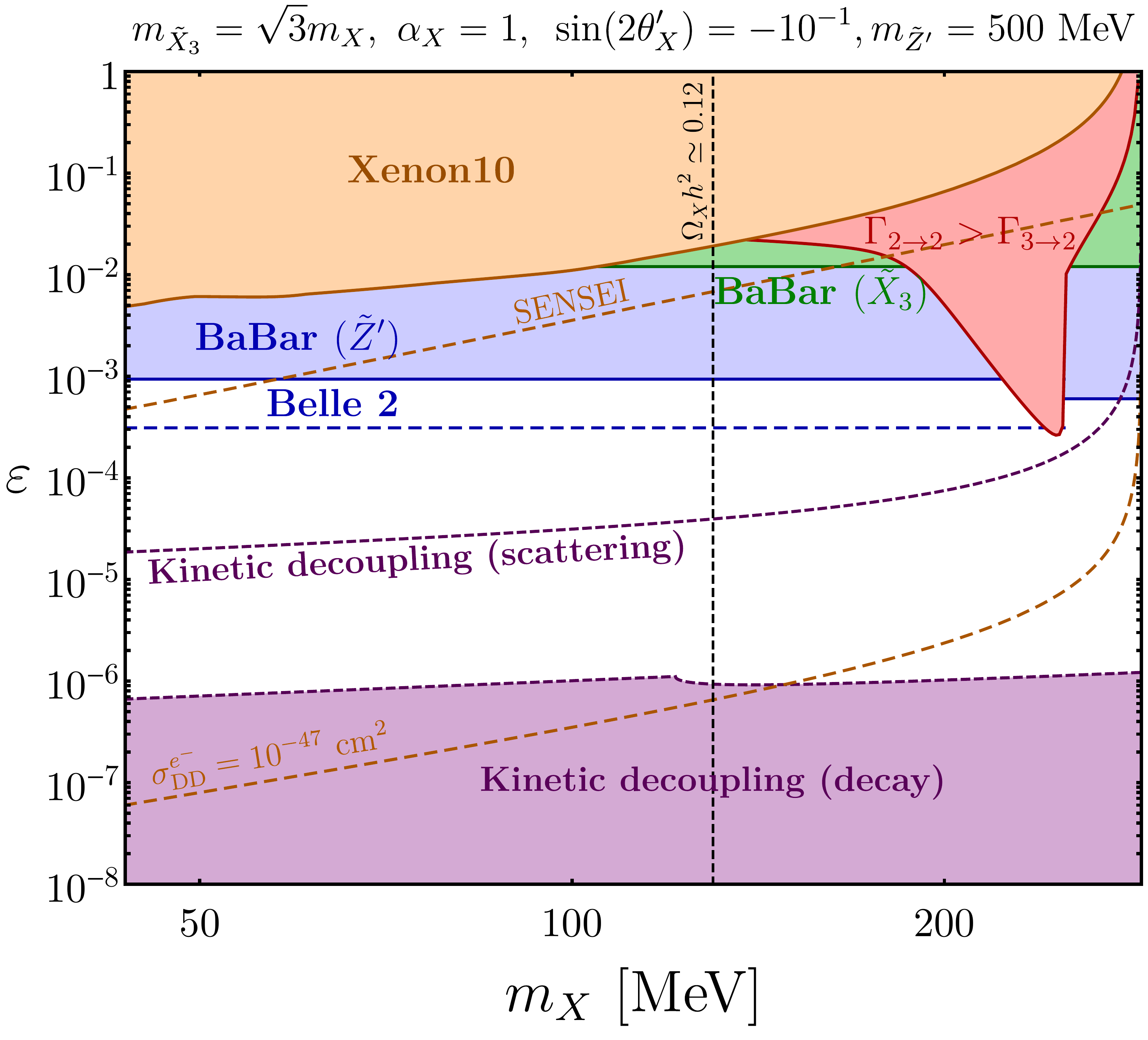}  \vspace{0.5cm} \\
                \includegraphics[height=0.34\textwidth]{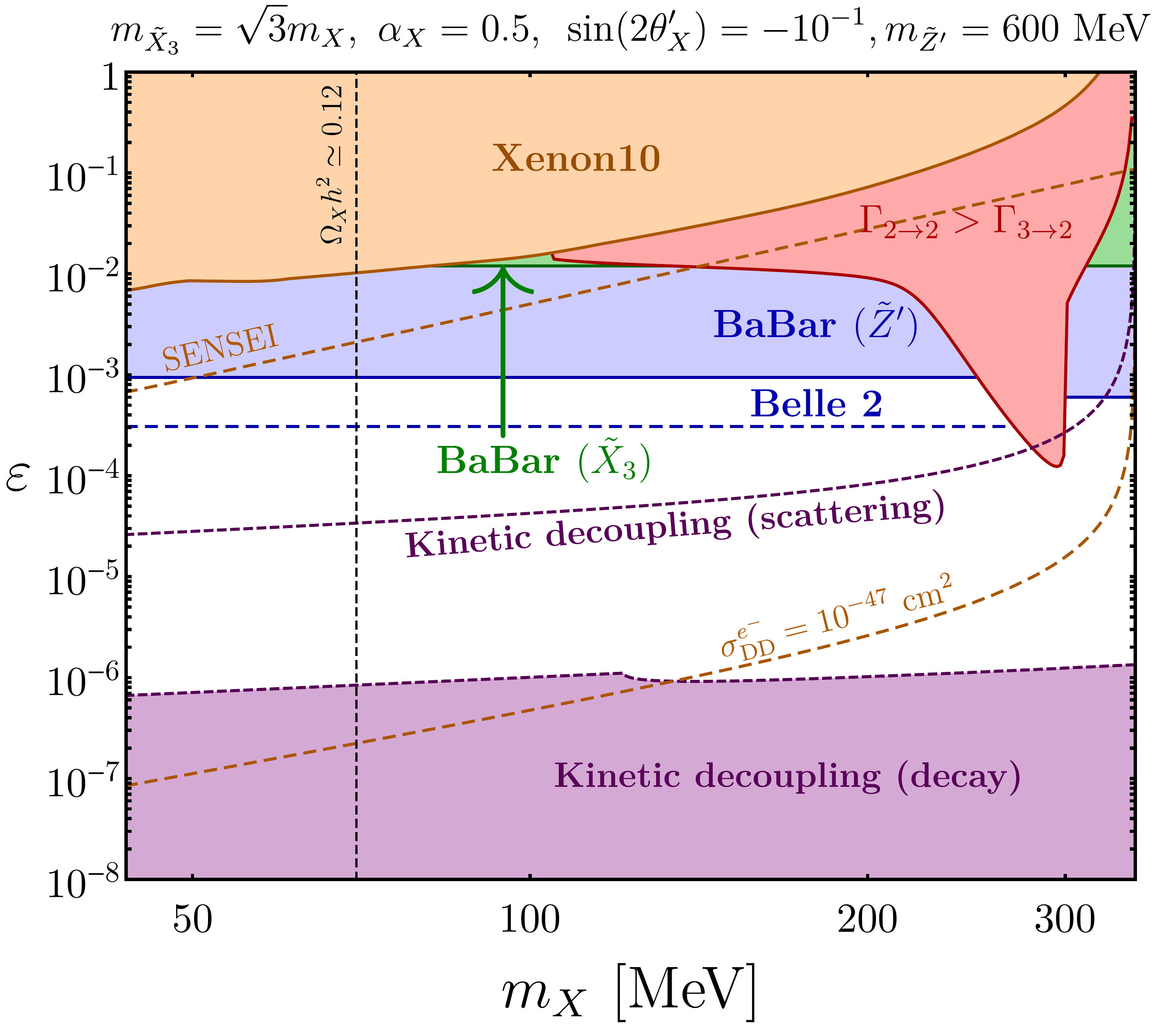} \quad
                    \includegraphics[height=0.34\textwidth]{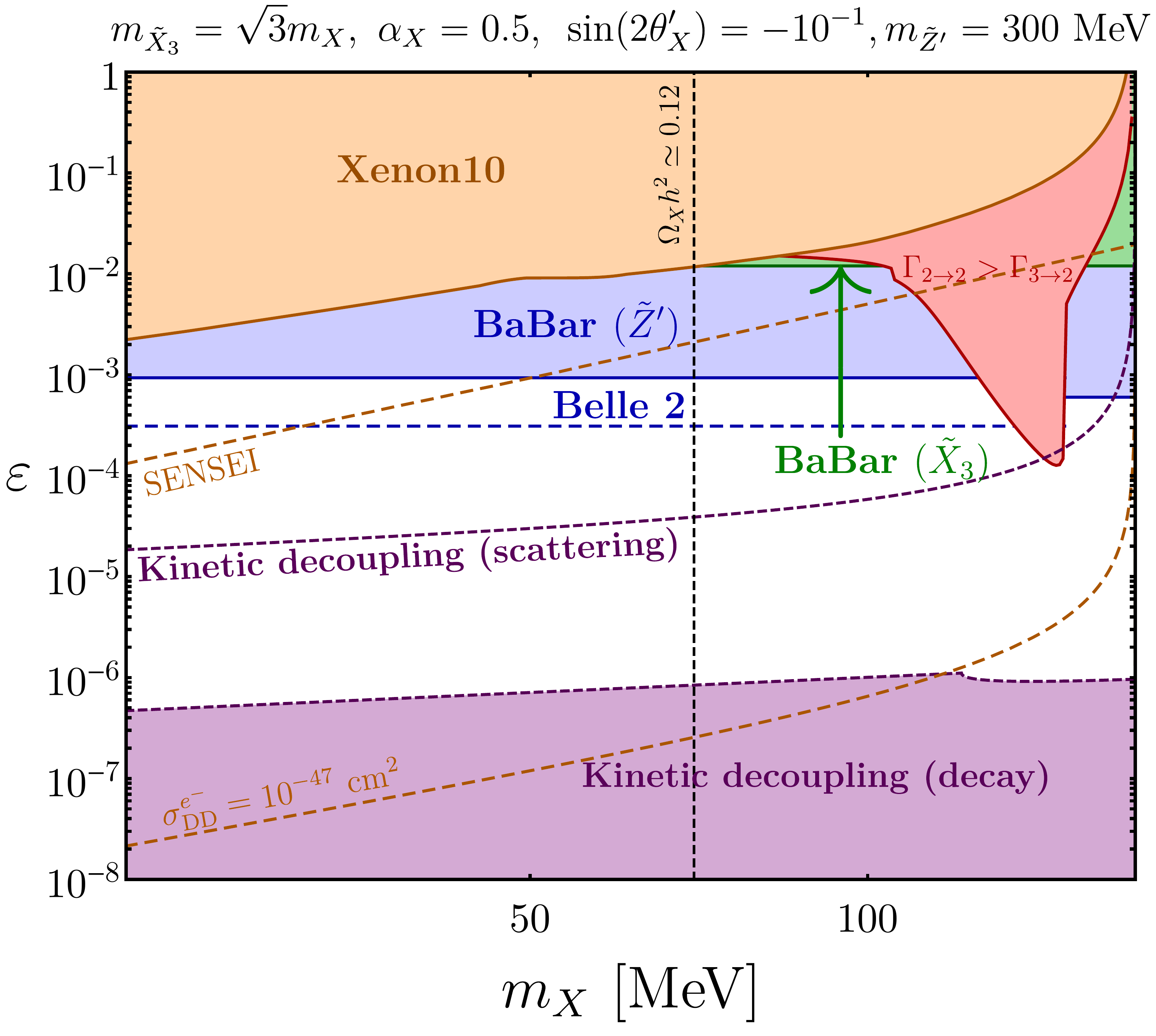} 
  \end{center}
  \caption{Various constraints on $\varepsilon$ vs $m_X$ for the case with quadruplet dark Higgs. The blue region and the region above the blue dashed line are excluded by invisible modes of ${\tilde Z}'$ in BaBar \cite{babar-inv} (observed) or Belle2  \cite{belle2-updated} (expected), respectively. The green region is excluded by visible modes of ${\tilde X}_3$ in BaBar \cite{babar-vis}  (dilepton$+$monophoton: observed).  Direct detection limit on DM-electron elastic scattering from XENON10 \cite{xenon10} (observed) and SENSEI-100 1yr  \cite{sensei} (expected) are shown in yellowish region and red dashed line, respectively, and contours with DM-electron elastic scattering cross section, $\sigma_{\rm DD}^{e^-}=10^{-47}\,{\rm cm}^2$, are also shown in orange dashed lines. Kinetic decoupling occurs in purple region and $2\rightarrow 2$ visible annihilations would be dominant in red region. For comparison, the DM-electron scattering process becomes negligible for kinetic equilibrium in the region below the purple dashed line. The relic density is saturated along the dashed black vertical lines. }
  \label{epsilon1}
\end{figure}

We consider the elastic scattering between dark matter and electron through $Z'$ portal couplings to achieve a kinetic equilibrium for SIMP dark matter. 
In this case, the corresponding momentum relaxation rate during freeze-out is given by
\bea
\gamma(T)\simeq 
\frac{31 \pi^3  e^2 \epsilon^2 g_X^2 \sin^2(2\theta'_X)T^6}{1512 m_X }\bigg(\frac{1}{m_{\tilde{X}_3}^2}-\frac{1}{m_{\tilde{Z}'}^2}\bigg)^2. 
\eea
Them, for kinetic equilibrium during SIMP freeze-out, we need to require
\bea
\gamma(T_\text{F})> H(T_\text{F}) \Big(\frac{m_X}{T_\text{F}} \Big)^2
\eea
where $T_\text{F}\simeq m_X/15$ is the freeze-out temperature and $H(T_\text{F})$ is the Hubble parameter at freeze-out.

On the other hand, since dark-neutral gauge boson ${\tilde X}_3$ has a similar mass as dark matter in our model, its abundance is not that suppressed around freeze-out temperature.
Thus, ${\tilde X}_3$ plays an important role of keeping dark matter in thermal equilibrium through the SM by the decays into electron or muon pairs, the kinetic equilibrium for dark matter can be also achieved by the elastic scattering between dark matter and ${\tilde X}_3$. Thus, in this case, we require
\be
n_{\tilde{X}_3}^{\rm eq} \Gamma_{\tilde{X}_3} > H(T_f)n_X^{\rm eq}.
\ee
We find that kinetic equilibrium for dark matter can be easily achieved by the elastic scattering between dark matter and dark-neutral gauge boson ${\tilde X}_3$ as far as the latter remains in kinetic equilibrium for a tiny gauge kinetic mixing $\epsilon\sim 10^{-6}$ in the parameter space of our interest. 

\begin{figure}
  \begin{center}
      \includegraphics[height=0.295\textwidth]{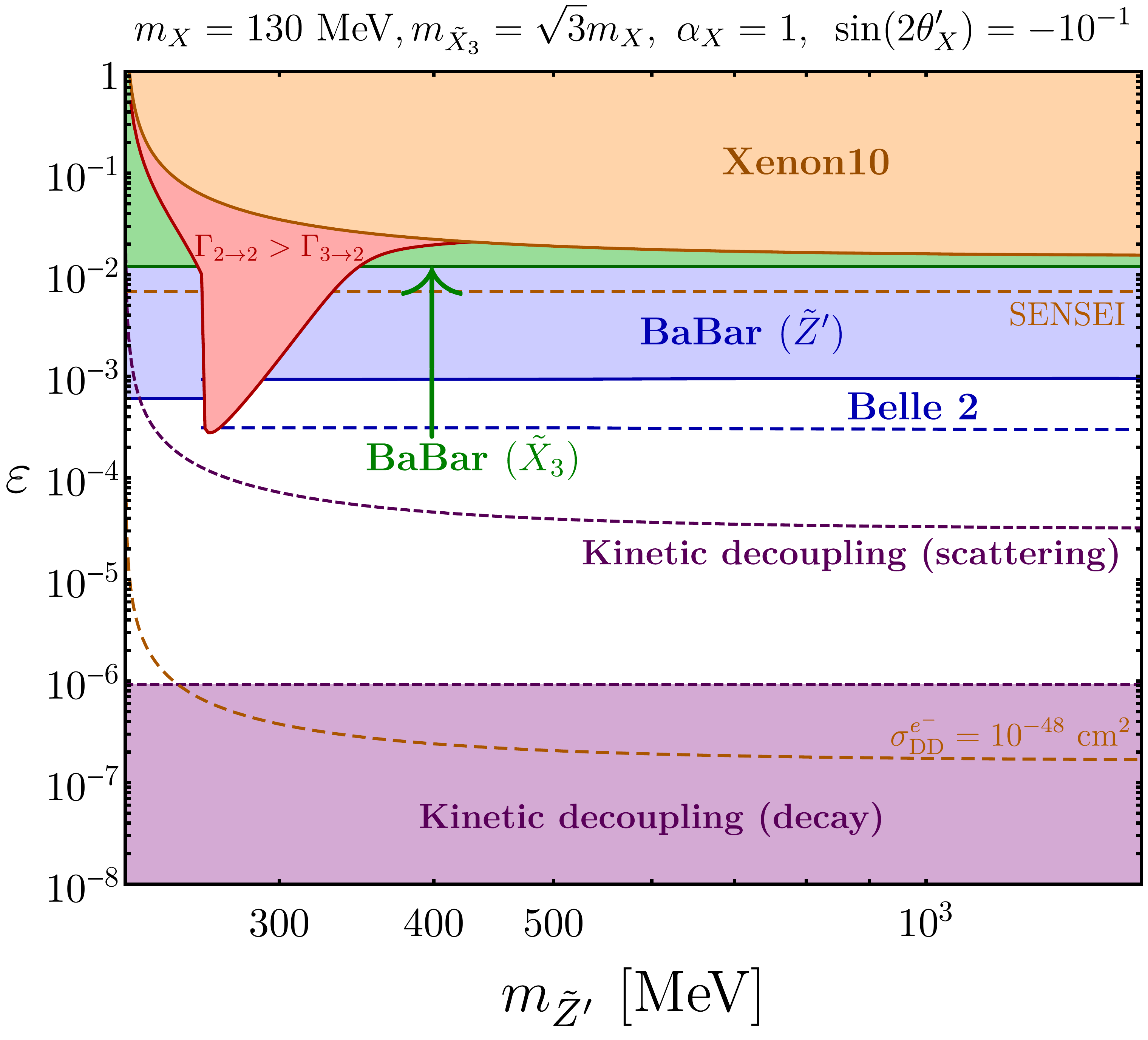}
            \includegraphics[height=0.295\textwidth]{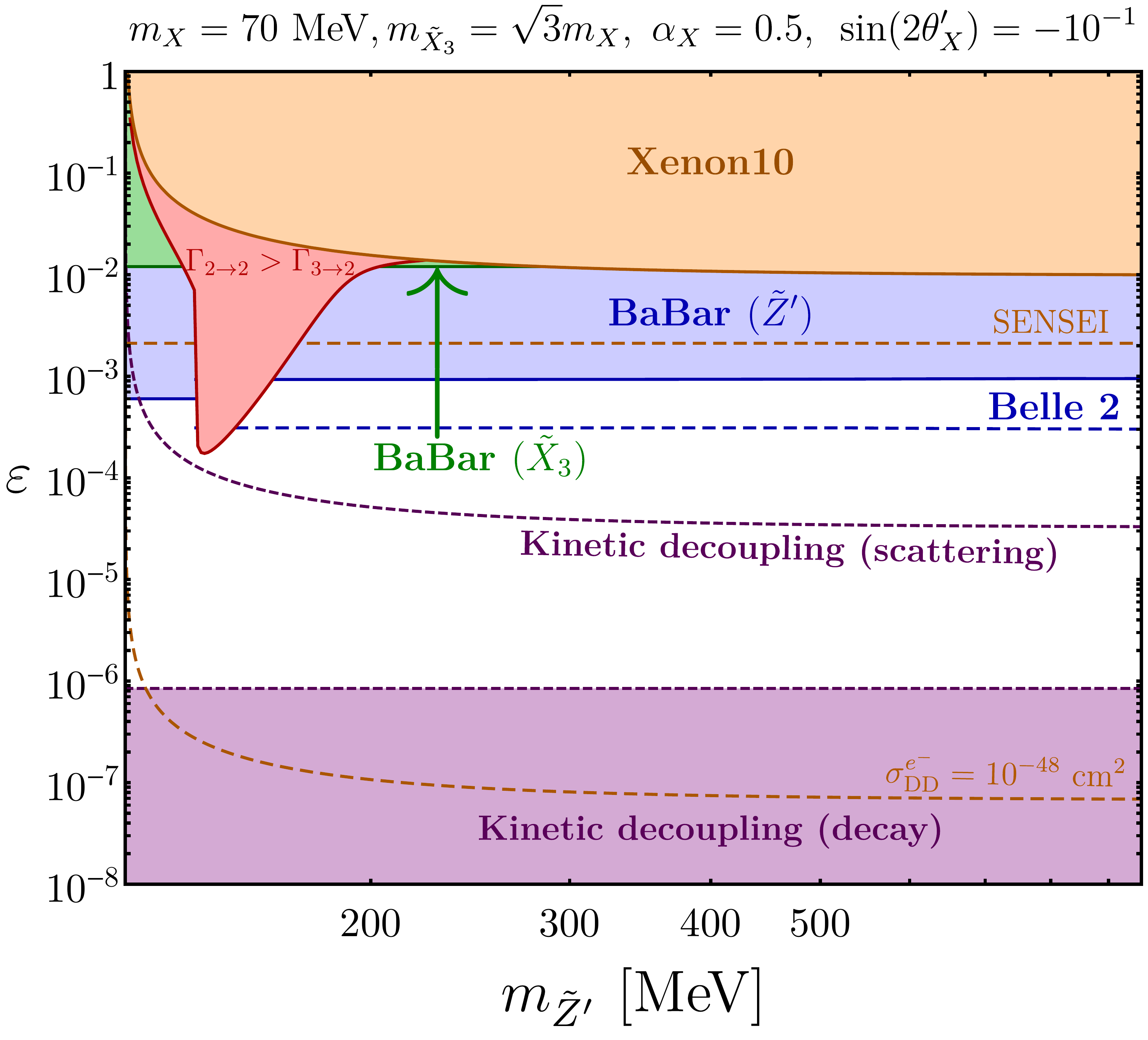}
               \includegraphics[height=0.295\textwidth]{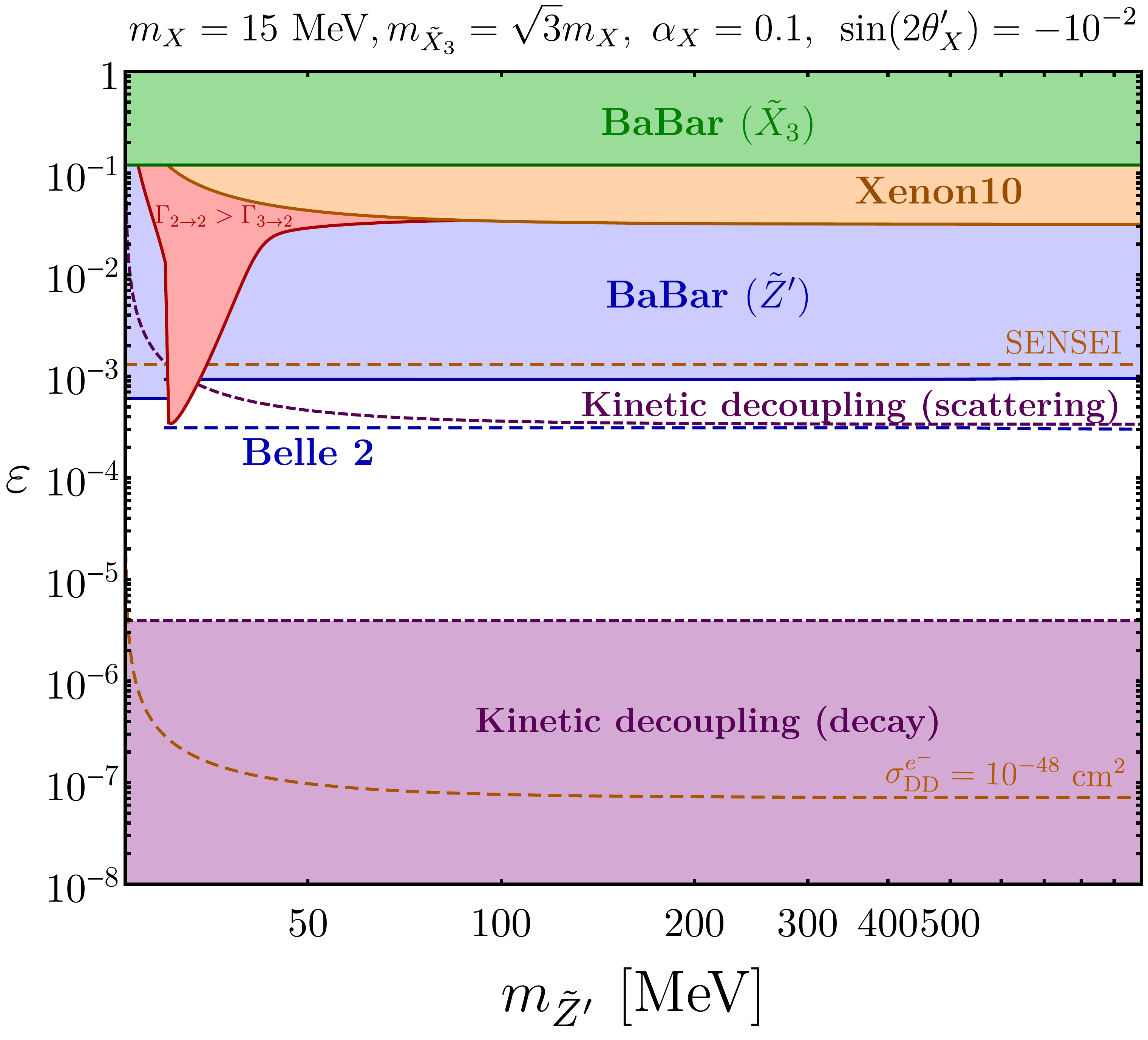} 
  \end{center}
  \caption{Similar constraints as in Fig.~\ref{epsilon1}, but in the parameter space for $\varepsilon$ vs $m_{{\tilde Z}'}$  for the case with quadruplet dark Higgs. }
  \label{epsilon2}
\end{figure}

In Fig.~\ref{epsilon1}, we showed the parameter space for $\varepsilon\approx c_W \xi$ vs $m_X$ with various constraints coming from the model consistency and experiments. We have assumed that the VEV of a quadruplet dark Higgs determines $SU(2)_X$ gauge boson masses and chosen different values for the DM self-coupling $\alpha_X=1, 0.5$ in the top and bottom panels, respectively, varying $m_{{\tilde Z}'}$.  We have similar plots for $\varepsilon$ vs  $m_{{\tilde Z}'}$ in Fig.~\ref{epsilon2}.

As for the theoretical constraints, dark matter becomes out of kinetic equilibrium already during freeze-out in purple region and $2\rightarrow 2$ visible annihilations such as $X_+ X_-\rightarrow {\tilde X}_3 {\tilde X}_3$ would be dominant in red region. The relic density is saturated along the black dashed vertical line in Fig.~\ref{epsilon1}, which is determined mostly by the SIMP processes. In Fig.~\ref{epsilon2}, we have chosen the values of $m_X$ and $\alpha_X$ such that the correct relic density is saturated.

We now turn to the experimental constraints on the model. As ${\tilde Z}'$ decays mostly invisibly into a dark matter pair, the observed limit from BaBar \cite{babar-inv} and the future projection  \cite{belle2-updated} from Belle2 \footnote{See also Ref.~\cite{belle2-inv} for the previous estimates on Belle2 sensitivity.}  can rule out the parameter space in blue region and in the region above the blue dashed line. On the other hand, ${\tilde X}_3$ decays visibly into a lepton pair in the SM so there are constraints from the monphoton$+$dilepton searches in BaBar \cite{babar-vis}, but they are not as strong as the invisible searches for ${\tilde Z}'$ in BaBar. Direct detection from electron-recoil signals in XENON10 \cite{xenon10} has excluded the yellowish region, sometimes being comparable than or even stronger than the visible bound from BaBar. The future updated  SENSEI-100 1yr  \cite{sensei} can reach the limit for light ${\tilde Z'}$ mediator, even beyond the Belle2 projection. For comparison, we also showed the contours with DM-electron elastic scattering cross section, $\sigma_{\rm DD}^{e^-}=10^{-47}\,{\rm cm}^2$, in orange dashed lines.

\subsection{SM $2\rightarrow 2$ annihilations and direct detection}

Light dark matter can also annihilate into a pair of leptons or mesons for $m_X\lesssim 1\,{\rm GeV}$. In particular, the annihilation cross section for $X_+ X_-\rightarrow f{\bar f}$, with $f$ being electron or muon, is before thermal average, 
\bea
(\sigma v_{\rm rel})_{X_+X_-\rightarrow f{\bar f}}= \frac{e^2 \epsilon^2 g_X^2 \sin^2(2\theta'_X) (m_f^2+2m_X^2)(m_{\tilde{X}_3}^2-m_{\tilde{Z}'}^2)^2}{16\pi (4m^2_X-m_{\tilde{X}_3}^2)^2(4m_X^2-m_{\tilde{Z}'}^2)^2 }\sqrt{1-\frac{m_f^2}{m_X^2}}\, v_{\rm rel}^2 .
\eea
The thermal average of the above annihilation cross section needs care near the resonances, except which, the DM $2\rightarrow 2$ annihilations are velocity-suppressed so they are not constrained by indirect detection experiments \cite{z3model}.

The ${\tilde X}_3$ and ${\tilde Z}'$ couple to both dark matter and electron so direct detection for light dark matter is relevant.
For $m_e, m_X , m_{{\tilde Z}'}, m_{{\tilde X}_3}\gg  p_X\simeq m_X v_{\rm X}$, the DM-electron elastic scattering cross section is approximately given by
\bea
\sigma_{\rm DD}^{e^-}&=& \frac{e^2\epsilon^2g_X^2 \sin^2(2\theta_X')m_e^2m_X^2}{4\pi (m_e+m_X)^2}\bigg(\frac{1}{m_{\tilde{X}_3}^2}-\frac{1}{m_{\tilde{Z}'}^2}\bigg)^2 \nonumber \\
&\approx& 9.3\times 10^{-42}\,{\rm cm}^2\, \left(\frac{\varepsilon}{10^{-4}} \right)^2\Big(\frac{\alpha_X}{1}\Big) \left(\frac{\sin(2\theta_X')}{0.1} \right)^2 \left(\frac{100\,{\rm MeV}}{m_{\tilde{X}_3}} \right)^2 \left(1-\frac{m_{\tilde{X}_3}^2}{m_{\tilde{Z}'}^2}\right)^2.
\eea
Thus, there are two mediators contributing to the DM $2\rightarrow 2$ annihilation and the DM-electron elastic scattering in our case, in particular, we can impose the direct detection constraints on the $Z'$ portal coupling with a single mediator by identifying  $1/m^4_{Z'}$ with $(1/m_{\tilde{X}_3}^2-1/m_{\tilde{Z}'}^2)^2$ as we have discussed for Figs.~\ref{epsilon1} and \ref{epsilon2}.

\section{Conclusions}

We have proposed a new model for Vector SIMP dark matter in the context of dark $SU(2)_X\times U(1)_{Z'}$ gauge theory with a similarity to the SM counterpart.  The mass splitting between dark matter and neutral components of $SU(2)_X$ are predicted by an approximate custodial symmetry in the dark Higgs sector, playing a crucial role for the production mechanism of dark matter due to self-interactions. The kinetic equilibrium for VSIMP is maintained during the freeze-out thanks to a gauge kinetic mixing between $U(1)_{Z'}$ and the SM, providing a testing ground for searches for light  mediators of order GeV scale or below at current and future collider and direct detection experiments.

\section*{Acknowledgments}

The work of YM was also supported by the France-US PICS MicroDark.  The work of HML and SMC is supported in part by Basic Science Research Program through the National Research Foundation of Korea (NRF) funded by the Ministry of Education, Science and Technology (NRF-2016R1A2B4008759, NRF-2018R1A4A1025334 and NRF-2019R1A2C2003738). 
The work of SMC is supported in part by TJ Park Science Fellowship of POSCO TJ Park Foundation. The work of MP was supported by the Spanish Agencia Estatal de Investigaci\'{o}n through the grants FPA2015-65929-P (MINECO/FEDER, UE), IFT Centro de Excelencia Severo Ochoa SEV-2016-0597, and Red Consolider MultiDark FPA2017-90566-REDC.
This research has also been supported by the (Indo-French) CEFIPRA/IFCPAR Project No.~5404-2. Support from CNRS LIA-THEP and the
INFRE-HEPNET of CEFIPRA/IFCPAR is also acknowledged.  
YM acknowledges partial support from the European Union Horizon 2020 research and innovation programme under the Marie Sklodowska-Curie: RISE InvisiblesPlus (grant agreement No 690575) and the ITN Elusives (grant agreement No 674896).

\def\theequation{A.\arabic{equation}}

\setcounter{equation}{0}

\vskip0.8cm
\noindent
{\Large \bf Appendix A: General dark gauge boson masses} 
\vskip0.4cm
\noindent

We discuss general gauge boson masses in the dark sector in our model.
In the presence of a dark Higgs field $F$  in the representation with $SU(2)_X$ isospin $I$, general dark gauge boson mass terms are given \cite{triplet} by
\bea
{\cal L}_{\rm SSB} &=& F^\dagger \Big(g^2_X[I(I+1)-(I_3)^2]X^\dagger_\mu X^\mu+g^2_X (I_3)^2 X_{3\mu} X^\mu_3  \nonumber \\
&&\quad+g^2_{Z'} (Z^{\prime })^2 Z'_\mu Z^{\prime \mu}+2 g_X g_{Z'}  (Z^{\prime }I_3)   Z'_\mu X^\mu_3  \Big) F.
\eea
For a singlet $S$, a doublet $\Phi$,  a triplet $T$, a quadruplet $Q_4$, a quintuplet $Q_5$, etc,  with $I=\frac{1}{2}, 1, \frac{3}{2}, 2$, etc,  respectively, we assign $Z'=q_S$ and $Z'=I_3=\frac{1}{2},1, \frac{3}{2}, 2$, etc,  resulting in $I(I+1)-(I_3)^2=I=\frac{1}{2}, 1, \frac{3}{2}, 2$, etc. 
Then, for nonzero VEVs with $S=\frac{1}{\sqrt{2}}\,v_S$, $\langle\Phi\rangle=\frac{1}{\sqrt{2}}\,(0,v_\Phi)^T$, and  and $T=(0, 0,\frac{1}{\sqrt{2}} v_T)^T$, $Q_4=(0, 0,0,\frac{1}{\sqrt{2}} v_{Q_4})^T$, $Q_5=(0, 0,0,0,\frac{1}{\sqrt{2}} v_{Q_5})^T$, etc, the dark-charged gauge bosons, $X_\mu, X^\dagger_\mu$, have masses, 
\be
m^2_X= \frac{1}{2} g^2_X \sum_I  I v^2_I \label{mass1}
\ee
where $v_I=v_\Phi, v_T, v_{Q_4}, v_{Q_5}$, etc.
On the other hand, the mass matrix for neutral gauge bosons, $Z'_\mu$ and $X_{3\mu}$, takes the following form,
\bea
M^2_{2\times 2}=m^2_{X_3}\left(\begin{array}{cc} \beta s^2_X  & -s_X c_X \\   -s_X c_X  & c^2_X  \end{array} \right) \label{mass22}
\eea
where $m^2_{X_3}\equiv (g^2_X+g^2_{Z'})\sum_I I^2 v^2_I$, 
$c_X\equiv \cos\theta_X$ and $s_X\equiv \sin\theta_X$, with $\sin\theta_X=g_{Z'}/\sqrt{g^2_X+g^2_{Z'}}$, and 
\bea
\beta\equiv 1+\frac{q^2_S v^2_S}{\sum_I I^2 v^2_I }. 
\eea
In the absence of the gauge kinetic mixing, the above mass matrix (\ref{mass22}) can be diagonalized by introducing a dark Weinberg angle as in the SM. 
Performing a rotation of dark gauge fields to mass eigenstates, ${\tilde Z}'_\mu, {\tilde X}_{3\mu}$,  as
\bea
\left(\begin{array}{c}  Z'_\mu \\ X_{3\mu} \end{array} \right)= \left(\begin{array}{cc} \cos\theta'_X & -\sin\theta'_X \\   \sin\theta'_X & \cos\theta'_X \end{array} \right) \left(\begin{array}{c}  {\tilde Z}'_\mu \\ {\tilde X}_{3\mu} \end{array} \right)
\eea
with
\bea
\tan(2\, \theta'_X)= \frac{2c_X s_X}{c^2_X-\beta\, s^2_X}, \label{gaugemix}
\eea
we obtain the mass eigenvalues for dark gauge bosons,
\bea
m^2_{ {\tilde Z}'} &=&m^2_{X_3}c^2_X  (1-\cot\theta'_X\, \tan\theta_X),  
\label{mass2} \\
m^2_{{\tilde X}_{3}} &=& m^2_{X_3}c^2_X(1+\tan\theta'_X\, \tan\theta_X). \label{mass3}
\eea
The results generalize our results in eqs.~(\ref{mass10})-(\ref{mass30}) with a singlet $S$ and a triplet $T$ in the main text.

For $\beta\gg 1$, from eq.~(\ref{gaugemix}), we get $\tan\theta'_X\approx -\frac{1}{\beta\tan\theta_X}$, leading to the approximate gauge boson masses,
\bea
m^2_{ {\tilde Z}'} &\approx& g^2_X\Big( \sum_I I^2 v^2_I\Big) \Big( 1+\beta \tan^2\theta_X\Big), \\
m^2_{{\tilde X}_{3}} &\approx& g^2_X \Big(\sum_I I^2 v^2_I\Big) \Big( 1-\frac{1}{\beta}\Big).
\eea
In this limit, the mass difference between the light gauge bosons is given by
\bea
m^2_{{\tilde X}_{3}} - m^2_X \approx g^2_X\sum_I I\Big(I-\frac{1}{2} \Big)v^2_I-\frac{1}{\beta}\, g^2_X\sum_I I^2 v^2_I. \label{genmassdiff}
\eea
Therefore, for $m^2_{{\tilde X}_{3}} >m^2_X$, we need $I>\frac{1}{2}$, namely, at least a triplet dark Higgs with nonzero VEV.

For instance, ignoring the mass splitting due to the dark Weinberg mixing and keeping only one Higgs representation with $I=\frac{1}{2}, 1, \frac{3}{2}, 2$, we get
\bea
m^2_{{\tilde X}_{3}} &\approx&  m^2_X, \quad 2m^2_X, \quad 3 m^2_X, \quad  4m^2_X.
\eea
Then, we get $\Delta\equiv (m_{{\tilde X}_{3}}-m_X)/m_X$ for $I=\frac{1}{2}, 1, \frac{3}{2}, 2$, as follows,
\bea
\Delta=0, \quad \sqrt{2}-1, \quad \sqrt{3}-1, \quad 1.
\eea
But, for general VEVs of all Higgs representations with $\frac{1}{2}\leq I\leq 2$, we can cover the entire range of the mass splitting continuously for $0\leq \Delta\leq 2$.

\def\theequation{B.\arabic{equation}}

\setcounter{equation}{0}

\vskip0.8cm
\noindent
{\Large \bf Appendix B: Dark Higgs masses} 
\vskip0.4cm
\noindent

In this appendix, we discuss general dark-charged and neutral Higgs boson masses in the presence of dark gauge symmetry breaking in our model and ensure that extra Higgs bosons can be safely decoupled in our consideration.

\underline{Triplet Higgs bosons}:

Minimizing the scalar potential for the dark triplet Higgs in eq.~(\ref{eq:Vhiggs}), we find that the VEV of the dark triplet Higgs  is related to the parameters in the scalar potential as follows,
\begin{equation}
    v_T = \sqrt{\frac{m^2_T}{\lambda_T-\tilde{\lambda}_T}}.
\end{equation}
Then, together with the Higgs portal coupling in eq.~(\ref{Higgsp}), dark Higgs bosons in the triplet receive masses,
\begin{equation}
 m_{h_T}^2=2v_T^2 (\lambda_T-\tilde{\lambda}_T)-\dfrac{1}{2}\lambda_{TH}v^2 \,, \quad 
 m_{h^{++}}^2=2v_T^2 \tilde{\lambda}_T-\dfrac{1}{2}\lambda_{TH}v^2,
\end{equation}
and there is a mixing between the neutral dark Higgs $h_T$ and the SM Higgs $h$ by
\be
\left( \begin{array}{c} h_1 \\ h_2 \end{array} \right)=  \left( \begin{array}{cc} \cos\theta & -\sin\theta \\  \sin\theta & \cos\theta \end{array} \right) \left( \begin{array}{c} h_T \\ h \end{array} \right)
\ee
where $h_{1,2}$ are mass eigenstates: $h_1$ is the triplet-like Higgs and $h_2$ the SM-doublet like Higgs. Then, the mass eigenvalues of neutral scalars are
\begin{align}
m^2_{h_{1,2}}=&\frac{1}{4} v^2 \Big(4 \lambda_H-\lambda _{TH}\Big) -\frac{1}{4}v_T^2 \Big(4 (\tilde{\lambda}_T-\lambda_T)+\lambda_{TH}\Big)\nonumber \\
&\mp \frac{1}{4}\Bigg[
\Big(v^2(4 \lambda_H-\lambda_{TH})-v^2_T(4 (\tilde{\lambda}_T-\lambda_T)+\lambda_{TH})\Big)^2 \nonumber \\
& \quad +4 v_T^2 \lambda_{TH} (3 v^2 \lambda_{TH} -4v_T^2 (\tilde{\lambda}_T-\lambda_T)  ) +16v^2 \lambda_H (v^2 \lambda_{TH}+4v_T^2(\tilde{\lambda}_T-\lambda_T) )
\Bigg]^{1/2},
\end{align}
and the mixing angle is
\be
\tan 2\theta =  -\frac{4 \, v \, v_T \lambda_{TH}}{ \lambda_{TH}v^2+4 \lambda_H v^2-v_T^2 (-4 \tilde{\lambda }_T+4 \lambda _T+\lambda _{TH})}.
\ee
In the limit of a small $\lambda _{TH}$, the mass eigenvalues of neutral scalars become
\begin{equation}
   m^2_{h_{1}}\approx 2v_T^2 \left(\lambda_T-\tilde{\lambda}_T\right), \quad m^2_{h_2}\approx 2\lambda_H v^2.
\end{equation}
Then, the vacuum stability bound, $\lambda_T>{\tilde \lambda}_T$, ensures the positive squared masses for neutral dark Higgs boson, $h_1\approx h_T$, for a small mixing quartic coupling.

In order to ensure the consistency of the dark vacuum, we require $m_{h^{++}}^2>0$ or
\begin{equation}
    \lambda_{TH}<\dfrac{4v_T^2 \tilde{\lambda}_T}{v^2}, \label{consistent}
\end{equation}
which naturally pushes the value of the quartic mixing to be $ \lambda_{TH} \lesssim 10^{-5}$ for $v_T \sim\text{ GeV}$, implying the mixing angle $\tan 2\theta \lesssim 10^{-5}$ to be below the bound from the Higgs  invisible decay.

Moreover, from the kinetic terms of the triplet complex field in eq.~(\ref{eq:Vhiggs}), we also derive the following interactions between dark Higgs and gauge bosons,
\begin{align}
    \mathcal{L}_\text{DH-DG}=&i h^{++} \overset{\leftrightarrow}{\partial}_\mu (h^{++})^\dagger \Big( g_X X_3^\mu+ g_{Z^\prime} Z^{\prime \mu} \Big) +g_X^2\Big(\dfrac{v_T+h_T }{\sqrt{2}} \Big) \Big( (h^{++})^\dagger X^\mu X_\mu+h^{++} X^\dagger_\mu X^{\dagger \mu} \Big) \nonumber\\
    &+|h^{++}|^2 \Big(  g_{Z^\prime}^2 Z^{\prime \mu} Z^{\prime}_{\mu}+2g_X g_{Z^\prime} Z^{\prime \mu}X_{3 \mu}+g_X^2 \left( X_3^\mu X_{3 \mu} +X^{\dagger \mu} X_\mu \right) \Big) \nonumber \\
    &+\Big(\dfrac{1}{2}h_T^2+v_T h_T \Big) \Big(  g_{Z^\prime}^2 Z^{\prime \mu} Z^{\prime}_{\mu}-2 g_X g_{Z^\prime} Z^{\prime \mu}X_{3 \mu}
    +g_X^2 \left( X_3^\mu X_{3 \mu} +X^{\dagger \mu} X_\mu \right) \Big).  \label{higgs-gauge}
\end{align}

\underline{Quadruplet Higgs bosons}:

For the quadruplet Higgs $Q_4 = \big(h^{(3)},h^{(2)},0,\frac{1}{\sqrt{2}}(v_{Q_4} + h_{Q_4})\big)^T$ with
\be
v_{Q_4}\ =\ \frac{2m_{Q_4}}{\sqrt{4\lambda_{Q_4} - 9\tilde{\lambda}_{Q_4}}},
\ee
the masses of dark-charged and neutral Higgs bosons are
\bea
m_{h^{(2)}}^2& =& 3 \tilde{\lambda}_{Q_4} v_{Q_4}^2 - \frac{1}{2} \lambda_{Q_4H}v^2\\
m_{h^{(3)}}^2&=& \frac{9}{2} \tilde{\lambda}_{Q_4} v_{Q_4}^2- \frac{1}{2} \lambda_{Q_4 H}v^2,\\
m_{h_{Q_4}}^2&=& \bigg(2\lambda_{Q_4} - \frac{9}{2}\tilde{\lambda}_{Q_4} \bigg)v_{Q_4}^2  - \frac{1}{2}\lambda_{Q_4 H}v^2 .
\eea
As a result, there are similar consistent conditions on the mixing quartic couplings for $m_{h^{(2)}}^2>0$ and $m_{h^{(2)}}^3>0$, as in eq.~(\ref{consistent}).
We can ignore the dark charged Higgs contributions in the later discussion, if they are heavy enough for $\lambda_{Q_4} v_{Q_4}^2\sim  \tilde{\lambda}_{Q_4} v_{Q_4}^2 \gg  m^2_X$.
Moerover, a similar vacuum stability bound, $\lambda_{Q_4}> \frac{9}{4}{\tilde \lambda}_{Q_4}$, ensures the positive squared masses for neutral dark Higgs boson $h_{Q_4}$, for a small mixing quartic coupling.

\underline{Quintuplet Higgs bosons}:

For the quintuplet Higgs $Q_5 = \big(h^{(4)},h^{(3)},h^{(2)},0,\frac{1}{\sqrt{2}}(v_{Q_5} + h_{Q_5})\big)^T$ with
\be
v_{Q_5}=\frac{m_{Q_5}}{\sqrt{\lambda_{Q_5} - 4\tilde{\lambda}_{Q_5}}},
\ee
 the masses of dark-charged and neutral Higgs are
\bea
m_{h^{(2)}}^2& =& 4 \tilde{\lambda}_{Q_5} v_{Q_5}^2 - \frac{1}{2} \lambda_{Q_5H}v^2,\\
m_{h^{(3)}}^2&=& 6\tilde{\lambda}_{Q_5} v_{Q_5}^2 - \frac{1}{2} \lambda_{Q_5 H}v^2,\\
m_{h^{(4)}}^2&= & 8 \tilde{\lambda}_{Q_5} v_{Q_5}^2 - \frac{1}{2} \lambda_{Q_5 H}v^2,\\
m_{h_{Q_5}}^2&=& 2\bigg(\lambda_{Q_5} - 4\tilde{\lambda}_{Q_5} \bigg)v_{Q_5}^2  - \frac{1}{2} \lambda_{Q_5H}v^2.
\eea
As a result, there are similar consistent conditions on the mixing quartic couplings for $m_{h^{(a)}}^2>0$ with $a=1,2,3$, as in eq.~(\ref{consistent}).
Similarly, he dark charged Higgs contributions can be neglected in the later discussion, when $ \lambda_{Q_5} v_{Q_5}^2\sim \tilde{\lambda}_{Q_5} v_{Q_5}^2\gg m^2_X$.
Moerover, a similar vacuum stability bound, $\lambda_{Q_5}> 4{\tilde \lambda}_{Q_5}$, ensures the positive squared masses for neutral dark Higgs boson $h_{Q_5}$, for a small mixing quartic coupling.

\end{document}